\documentclass[12pt]{JHEP3}
\usepackage{amsmath,amssymb,epsfig}

\relax

\title{From bubbles to foam: dilute to dense  evolution of hadronic wave function at high energy.}

\author{Alex Kovner${}^{\,1}$, Michael Lublinsky${}^{\,2}$ and 
Urs Achim Wiedemann${}^{\,3}$\\
\vspace{0.1in}

${}^{\,1}$Physics Department, University of Connecticut, 2152 Hillside road, Storrs, CT 06269, USA
\vspace{0.1in}

${}^{\,2}$Physics Department, State University of New York, Stony Brook, NY 11794, USA
\vspace{0.1in}

${}^{\,3}$Department of Physics, CERN, Theoretical Physics Unit, CH-1211 Geneva 23
\vspace{0.1in}

E-mail addresses: {\tt kovner@phys.uconn.edu, lublinsky@phys.uconn.edu, Urs.Wiedemann@cern.ch}
}

\abstract{We derive the evolution
of a hadronic light cone wave function with energy at weak coupling. Our derivation is valid
both in the high and the low partonic density limit, and thus
encompasses both the JIMWLK and the KLWMIJ evolution. The hadronic
wave function is shown to evolve by the action of the
Bogoliubov-type operator, which diagonalizes on the soft gluon sector 
the light-cone hamiltonian in the presence of an arbitrary valence charge density. 
We find explicitly the action of this
operator on the soft as well as the valence degrees of freedom of
the theory. 
}

\preprint{ CERN-PH-TH/2007-079}

\begin{document}
\def\vev#1{\langle#1\rangle}
\def\ov{\over}
\def\le{\left}
\def\ri{\right}
\def\ha{{1\over 2}}
\def\lam{{\lambda}}
\def\Lam{{\Lambda}}
\def\al{{\alpha}}
\def\ket#1{|#1\rangle}
\def\bra#1{\langle#1|}
\def\vev#1{\langle#1\rangle}
\def\det{{\rm det}}
\def\tr{{\rm tr}}
\def\Tr{{\rm Tr}}
\def\NN{{\cal N}}
\def\th{{\theta}}

\def\Om{{\Omega}}
\def \th{{\theta}}

\def \lam {\lambda}
\def \om {\omega}
\def \ra {\rightarrow}
\def \ga {\gamma}
\def\sig{{\sigma}}
\def\ep{{\epsilon}}
\def\apr{{\alpha'}}
\newcommand{\p}{\partial}
\def\LL{{\cal L}}
\def\TT{{\cal T}}
\def\CC{{\cal C}}

\def\tir{{\tilde r}}

\newcommand{\bea}{\begin{eqnarray}}
\newcommand{\eea}{\end{eqnarray}}
\newcommand{\nn}{\nonumber\\}
%%%%%%%%%%%%%%%%%%%%%%%%%%%%%%%%%%%%%%%%%%%%%%%%%%%%%%%%%%%%%%%%%%%%%%%%%%%%%%%%
\section{ Introduction.}
%%%%%%%%%%%%%%%%%%%%%%%%%%%%%%%%%%%%%%%%%%%%%%%%%%%%%%%%%%%%%%%%%%%%%%%%%%%%%%%%

The problem of calculating hadronic scattering amplitudes at high energy is an old one. 
It goes back to classical works of Gribov on Reggeon Field Theory \cite{gribov} in the pre-QCD days. Within the framework of QCD this question has been addressed from different points of view 
\cite{BFKL,GLR,Bartels,BKP}. 

In the last ten years or so the subject has seen new developments. Some of these have been triggered by Mueller's reformulation of the BFKL equation \cite{BFKL} in terms of the dipole model \cite{dipoles,NP} with additional input provided by the functional approach of \cite{mv}. The result was the derivation of the functional evolution equation for the hadronic amplitude - the so-called Balitsky hierarchy \cite{balitsky} or JIMWLK equation \cite{JIMWLK,cgc}; and its simplified mean field version due to Kovchegov \cite{kovchegov}. This evolution takes into account coherent emission effects in the dense hadronic wave function, or in other words partonic saturation effects. These effects lead to unitarization of the scattering amplitude. Although the language of this approach is different from the original Reggeon Field Theory, a direct relation between the JIMWLK evolution and the QCD Reggeon Field Theory has been investigated recently \cite{7P,yinyang}.

In the last couple of years spurred on by observations of \cite{shoshi}, the realization has emerged that the existing evolution equations, which are tailored to describe the situation when a small perturbative projectile scatters off a large dense target, do not include the so called Pomeron loop effects. The effort to account for the Pomeron loops using the probabilistic view of the evolution \cite{pomloops} has lead to interesting analogies between QCD and statistical systems \cite{stat}. Alternative approaches based on effective
Lagrangian can be found in \cite{Verlinde,LipatovFT,HIMST,Bal,Braunploop} 

Another avenue that has been explored in this context is the direct approach to the evolution of the hadronic wave function \cite{KW,klwmij,duality,remarks2}. This approach yielded the evolution equation valid in the limit opposite to that of JIMWLK, namely when the hadronic wave function is dilute. This so called KLWMIJ equation \cite{klwmij} is related to the JIMWLK equation by the dense-dilute duality transformation \cite{duality}. 
The basic strategy of this approach is to calculate the light cone hadronic wave function of soft gluonic  modes, given the color charge density $j^a(x)$ due to the 'valence' modes - the modes with large longitudinal momentum. When the hadron is boosted, the longitudinal momentum of the soft modes is increased and they contribute to the scattering matrix and other physical observables \cite{zakopane}. 
The evolution of any physical observable is therefore in principle completely determined once we know the soft part of the wave function. 

So far the hadronic wave function has been calculated only in the KLWMIJ limit, namely when the valence charge density is small; $j^a(x)\sim g$. The JIMWLK evolution on the other hand is valid when $j^a(x)\sim 1/g$, but no wave function evolution is available in the JIMWLK regime. The derivation of \cite{balitsky} is given directly for the scattering matrix. The original derivation of \cite{JIMWLK} is not far in spirit from the wave function form of the evolution, however it involves additional approximations which do not allow to read off the evolution of the wave function directly from the 
JIMWLK equation. 

The main motivation to know the explicit form of the wave function evolution comes from the possibility to use it to derive the generalization of the JIMWLK/KLWMIJ evolution that includes the Pomeron loops. 
The knowledge of the wave function is also crucial to be able to address a wide range of semi-inclusive observables \cite{KLW}.

In the present paper we derive the soft gluon wave function valid at any physically interesting value of the valence color charge density. The expression we derive is valid both in the JIWMLK and the KLWMIJ limits as well as at any value of the valence charge density which interpolates between the two: 
$g \leq j\leq 1/g$ .
We do this by diagonalizing the leading part of the light cone Hamiltonian on the soft gluon sector. The transformation that diagonalizes the Hamiltonian turns out to be of the Bogoliubov type with parameters depending on the valence color charge density operator. We find explicitly the action of this transformation on the basic quantum degrees of freedom: the soft components of the vector potential $A_i^a(x,x^-)$ and the valence color charge density $j^a(x)$.

We show that the expression for the wave function indeed reproduces the JIMWLK and the KLWMIJ evolution equations. To reproduce the KLWMIJ equation one simply neglects the coherent emission effects in the wave function.  Thus the evolution of the wave function in this limit is strictly perturbative\cite{klwmij}. The nontrivial physics in this limit is entirely due to the multiple scattering corrections in the scattering amplitude.
 On the other hand to derive the JIMWLK limit we keep all the coherent emission effects in the wave function. However as we show explicitly below, in this limit we neglect certain multiple scattering corrections in the scattering amplitude. Physically this is justified in the situation where our hadron scatters on a perturbatively small target, which is when the JIMWLK evolution is valid. 
 
 To derive the evolution equation which includes Pomeron loops exactly we have to keep both types of effects in the evolution of the scattering amplitude. Within the present framework this looks like a tractable problem. It is however beyond the scope of the present paper and is left for future work.
 
The paper is structured as follows. In Sec. II we recall the general framework of the high energy evolution. Secs. III and IV are the main part of this paper. Sec. III is devoted to the derivation of the "vacuum" wave function of the soft gluon Hilbert space in the presence of the valence color charge density.
In Sec. IV we show that this diagonalization is achieved by the action on the free vacuum of a Bogoliubov type operator and derive explicitly the action of this operator on the soft and valence degrees of freedom.
 In Sec.V we show how both the JIMWLK and the KLWMIJ evolution equations follow from the wave function we have found in Sec. III in different limits. Finally a discussion is presented in Sec. VI.

%%%%%%%%%%%%%%%%%%%%%%%%%%%%%%%%%%%%%%%%%%%%%%%%
\section{High energy evolution}

The logic of our approach is the same as described in \cite{klwmij,zakopane}. Suppose that at some 
initial rapidity $Y_0$ we know the wave function of a hadron. In the gluon Fock space it has a generic 
form  (we work in the $A^-=0$ gauge)
\begin{equation}
|P\rangle_{Y_0}\,=\,\Psi[a^{\dagger a}(x,k^+)]\,|0\rangle \ \ .
\end{equation}
There is some minimal longitudinal momentum $k^+=\Lambda$ below which there are no
gluons in this wave function. More precisely, the number of soft gluons with $k^+\leq \Lambda$ is not zero but is perturbatively small so that their contribution to the scattering amplitude at $Y_0$ is a small perturbative correction and can be neglected.

We are interested in describing the scattering of this hadron on some target. The target is described by some distribution of color fields $\alpha_T\equiv A^+$ with a probability density distribution
$W_T[\alpha_T]$.
The second-quantized $S$-matrix operator in the eikonal approximation 
(in which we are working throughout this paper) is given by
\begin{equation}\label{s1}
\hat S\,=\,\exp\left[i\int d^2x\, j^a(x)\,\alpha^a_T(x)\right]\, ,
\end{equation}
where 
\begin{equation}
 j^a(x)\,=\,g\,\int_{k^+>\Lambda} \frac{dk^+}{2\pi}\,a^{\dagger b}(x,k^+)\,T^a_{bc}\,a^c(x,k^+)
\end{equation}
is the color charge density operator at the transverse position $x$
( with $T^a_{bc}=if^{abc}$ - the generator of the color group in the adjoint representation).
After scattering on a particular configuration of the target field the hadronic wave function 
becomes
\begin{equation}
\hat S|P\rangle_{Y_0}\,=\,\Psi[S^{ab}(x)\,a^{\dagger b}(x,k^+)]\,|0\rangle\, ,
\end{equation}
where $S^{ab}(x)$ is a unitary matrix - the single gluon scattering matrix.
Since the scattering amplitude is sensitive only to the color charge density in the hadronic wave
function and not to any other characteristic of the hadron, we can think of this wave function as being specified by some distribution
of $j^a(x)$. The
color charge density correlators are determined in terms of the
'probability density functional' $W[j]$ via
\begin{equation}
\label{W}
 \langle j^{a_1}(x_1)...j^{a_n}(x_n)\rangle_{Y_0}\,=\,\int
Dj\,W_{Y_0}[j]\,j^{a_1}(x_1)...j^{a_n}(x_n)\, .
\end{equation}
The forward scattering amplitude is then given by 
\begin{equation}
{\cal S}\,=\,\,\int\, D\alpha_T^{a}\,\, W^T_{Y-Y_0}[\alpha_T(x)]\,\,\Sigma^P_{Y_0}[\alpha_T(x)]\, ,
\label{ss}
\end{equation}
where
\begin{equation}\label{sigma}
\Sigma^P_{Y_0}[\alpha_T]\,\,=\,
\langle 0|\Psi^*[a(x,k^+)]\,\hat S\,\Psi[a^\dagger(x,k^+)]|0\rangle=\,\,\int Dj\,\,W_{Y_0}[j]\,\,
\,\exp\left[i\,\int d^2x\,j^a(x)\,\alpha_T^a(x)\right]\,.
% \label{s}
\end{equation}
The total rapidity of the process is $Y$ while the target is assumed to be evolved to rapidity $Y-Y_0$.
Here, $W^T$ characterizes the distribution of color fields $\alpha_T$ in the target, while
$W\left[j\right]$ characterizes the distribution of color charges in the projectile. Due to Lorentz
invariance $\cal S$ is $Y_0$ independent.

The evolution of  the $S$-matrix (\ref{ss}) with energy in  the high energy
limit has the generic form
\begin{equation}
-\frac{d}{d\,Y}\,{\cal S}\,=\,\int\, D\alpha_T^{a}\,\, W^T_{Y-Y_0}[\alpha_T(x)]\,\,\,
H^{RFT}\left[\alpha_T,\frac{\delta}{\delta\,\alpha_T}\right]\,\,\,
\Sigma^P_{Y_0}[\alpha_T(x)]\, ,
\label{hee}
\end{equation}
where $H^{RFT}$ is the Hermitian kernel of high energy evolution, which can be viewed as acting
either to the right or to the left:
\begin{equation}
-{\partial\over\partial Y}\,\Sigma^P\,\,=\,\,H^{RFT}\left[\alpha_T,\,{\delta\over\delta\alpha_T}\right]\,\,\Sigma^P[\alpha_T]\,;
\ \ \ \ \ \ \ \ \ 
-{\partial\over\partial Y}\,W^T\,\,=\,\,
H^{RFT}\left[\alpha_T,\,{\delta\over\delta\alpha_T}\right]\,\,W^T[\alpha_T]\,.
\label{dsigma}
\end{equation} 

The color charge density operators are the generators of the 
$SU(N_c)$ algebra and as such do not commute
$$
[j^a(x),\ j^b(y)]\ =\ i\, f^{abc}\,j^c(x)\,\delta^2(x-y) \, .
$$
As explained in detail in \cite{klwmij}, to properly take into
account the non commuting nature of the charge density operators
$j(x)$ and to still be able to represent wave function averages in terms of the functional integral over 'classical' fields $j^a$, one has to assign to $j$ an additional 'longitudinal'
coordinate. Thus in effect $j^a(x)\rightarrow j^a(x,x^-)$, where the value of $x^-$ simply
keeps track of the order of the operators $j$ in the correlation
function eq.(\ref{W}). An analogous `longitudinal coordinate' should be assigned to the target field $\alpha_T$. Since in this paper we work in the Hamiltonian formalism and explicitly keep track of the commutation relations of the quantum operators $j^a(x)$, we will not need to dwell on this additional longitudinal coordinate. 

The preceding discussion is given in the situation when the increase of rapidity is assigned to the target. One can equally well boost the projectile. The evolution of the projectile probability density functional $W[j]$ is related to that of $\Sigma[\alpha_T]$ since the two are related by the functional Fourier transform  eq.(\ref{sigma})
\begin{equation}
-{\partial\over\partial Y}\,W[j],=\,\,H^{RFT}\left[{\delta\over \delta j},\,-j\right]\,\,W[j]\, .
\end{equation}

As the hadron is boosted by rapidity $\Delta Y$, the longitudinal momenta of the gluons in its wave function are scaled by the boost parameter $k^+\rightarrow e^{\Delta Y}k^+$. Thus some gluons in the wave function emerge after boost with the longitudinal momenta above the cutoff $\Lambda$ and have to be taken into account in the calculation of the scattering amplitude. The number of thus 'produced' additional gluons in the wave function is proportional to the total longitudinal phase space $\int \frac{dk^+}{k^+}=\Delta Y$. 

To find the evolution of the scattering amplitude we need two ingredients. First we have to solve for the initial hadronic wave function with greater accuracy on the soft gluon Hilbert space than is necessary to calculate the scattering amplitude at the initial rapidity $Y_0$. Second we need to take into account the contribution of these soft gluons into the scattering amplitude at the rapidity $Y=Y_0+\Delta Y$, which amounts to the transformation
\begin{equation}
j^a(x)\,\rightarrow \,j^a(x)\,+\,j^a_{soft}(x)\, , \ \ \ \ \ \ \ \ \ \ \ \ 
j^a_{soft}(x)\,=\,g\,\int_{\Lambda \,e^{-\Delta Y}}^{\Lambda}\frac{dk^+}{2\pi}a^{\dagger b}(x,k^+)\,
T^a_{bc}\,a^c(x,k^+)
\end{equation}
in eq.(\ref{s1}).
This transformation is conveniently represented in terms of the charge density shift operator (which also has the meaning of the 'dual' to the Wilson line operator\cite{duality})
\begin{equation}
\hat R_a\,\,=\,\,
\exp\left[{\int d^2 z\,j^c_{soft}(z)\,{\delta\over\delta j^c(z)}}\right]\, ,
\ \ \ \ \ \ \ \ \ \ \ \ \ \ \ \ \ \ \ \ \ \ \ 
j^c(x)\rightarrow \hat R_a\,j^c(x)\, .
\label{rr}
\end{equation}

 The crucial part of this program is the knowledge of the wave function on the soft gluon part of the Hilbert space, $k^+\leq \Lambda$ with some minimal accuracy.
The calculation of this wave function is
the subject of the next section.

 The QCD light cone Hamiltonian $H$ 
responsible for the dynamics of the soft modes is diagonalized by the 
action of a unitary operator $\Omega_{\Delta Y}$, where $\Delta Y$ corresponds to the phase space
volume  occupied  by the soft modes.
Equivalently, the vacuum wave function of the soft modes in the presence of the valence color charges  is
$\Omega\,|P\rangle$. The kernel of the high energy evolution, $H^{RFT}$ is related to $\Omega$ as \cite{yinyang}:
\begin{equation}\label{chi}
H^{RFT}\,=-\,\lim_{\Delta Y\rightarrow 0}{\langle 0_a|\Omega^\dagger_{\Delta Y}(j,a)\,
\left(\hat R_a\,-\,1\right)\,\Omega_{\Delta Y}(j, a)|0_a\rangle\over \Delta Y}\,.
\end{equation}
 We will find below that  $\Omega$ is an operator of the
Bogoliubov type for any physically interesting $j$:
$$
\Omega\,=\,{\cal C}\,{\cal B}
$$
with ${\cal C}$ denoting a coherent operator, which is the exponential of an operator linear in the soft fields $A$, whereas
${\cal B}$ is an exponential of an operator quadratic in $A$. In the dilute limit $j\sim g$ we have ${\cal B}=1$ and the coherent operator $\cal C$ leads to the KLWMIJ evolution\cite{klwmij}. For dense
systems $j\sim 1/g$, the Bogoliubov operator ${\cal B}$ also contributes to the leading order evolution kernel $H^{RFT}$. 
We derive
the action of $\Omega$ on both the valence and soft degrees of freedom,  
which enter equation (\ref{chi}).  The JIMWLK Hamiltonian \cite{JIMWLK} is obtained  from the general expression (\ref{chi}) in the limit of weak target fields $\alpha_T$ expanding $\hat R_a$ to
second order in $\delta/\delta j$.

%%%%%%%%%%%%%%%%%%%%%%%%%%%%%%%%%%%%%%%%%%%%%%%%%%%
\section{Diagonalizing The Soft Gluon Hamiltonian}
We will proceed in the following steps.
 In section~\ref{sec2.1} we formulate the light-cone Hamiltonian for soft gluon modes
$k^+ < \Lambda$, coupled to the  color charge
density of the hard modes $k^+ > \Lambda$. We observe that the zero modes of the vector potential are not independent degrees of
freedom, but are constraint by the residual gauge fixing and the requirement of finiteness of energy. In
section~\ref{sec2.2}, we solve the resulting constraints. In section~\ref{sec2.3}, 
we  diagonalize the resulting Hamiltonian, by first finding the 
complete set of solutions to the classical equations of motion, and then expanding the field
operators in this basis. To ensure the canonical commutation relations for the creation and annihilation operators associated with these basis functions, a proper normalization of
the classical solutions is needed. This normalization is found in section~\ref{sec2.4}. 
\subsection{The Hamiltonian and the canonical structure}
\label{sec2.1}
 The starting point of our approach is the light cone
hamiltonian of QCD \cite{Brodsky}
\begin{eqnarray}\label{hamiltonian}
  H = \int_{k^+>0} \frac{dk^+}{2\pi}\, d^2 x\,
  \left( \frac{1}{2} \Pi_a^-(k^+,x)\, \Pi_a^-(-k^+, x)
      + \frac{1}{4} G_a^{ij}(k^+, x)\, G_a^{ij}(-k^+, x)
      \right)\, ,
\end{eqnarray}
where the electric and magnetic pieces have the form
\begin{eqnarray}
  \Pi_a^-(x^-,{ x}) &=& -\frac{1}{\partial^+}
       \left( {D^i} \partial^+A_i\right)^a(x^-, x)\, ,
       \nonumber \\
  G_a^{\mu\nu}(x^-, x) &=&
  \partial^\mu A_\nu^a(x^-,x) -  \partial^\nu A_\mu^a(x^-, x)
     - g f^{abc}\, A^b_{\mu}(x^-, x)\, A^c_{\nu}(x^-, x)\, ,
\end{eqnarray}
and the covariant derivative is defined as
\begin{equation}
{D}_i^{ab}\ \Phi^b\,=\,\left(\partial_i\,\delta^{ab}\ -\ g\,f^{acb}\,A_i^c\right)\ \Phi^b\, .
\end{equation}

Our aim is to diagonalize this Hamiltonian on the Hilbert space of
soft gluon modes - those with longitudinal momenta smaller than some scale $\Lambda$. We
assume that the valence part of the wave function (the component
of the full wave function which does not contain soft modes) is
known and is completely specified by the correlation function of the
color charge density
\begin{equation}
j^a(x)\equiv igf^{abc}\int_{k^+>\Lambda}\frac{dk^+}{2\pi}\,a^{b\,\dagger}_i(k^+,x)\,a^c_i(k^+,x)\, .
\end{equation}
The soft modes are the interesting dynamical degrees of freedom of our problem, and they
interact with the valence ones via eikonal coupling in the
Hamiltonian. The Hamiltonian for the soft modes is then given by
eq.(\ref{hamiltonian}) with the substitution
\begin{eqnarray}
  \Pi_a^-(k^+, x) &=& \frac{1}{i(k^++i\epsilon)}
  \partial^i \partial^+A^a_i(k^+,x)
  + \frac{1}{-i(k^++i\epsilon)} j^a( x)
  \nonumber \\
  && + g \frac{1}{-i(k^++i\epsilon)} \,f^{abc}\,
  \int_{|p^+|<\Lambda} \frac{dp^+}{2\pi}
  A^b_i(k^+-p^+,x)\, (-ip^+)\, A^c_i(p^+, x)\, .
\end{eqnarray}
The soft fields $A$ are defined only below the longitudinal momentum cutoff $\Lambda$, but we will not
explicitly indicate it in the following.

 The canonical structure of the
theory is determined by the commutation relations of the fields. As
we will see, the zero momentum mode of the field $A$ is non dynamical
and is determined by the residual gauge fixing (still not specified so far on top of the usual light cone gauge condition $A^+=0$) and the constraint of
finiteness of energy. We denote by $\tilde A$  the part of the field
that does not contain the mode with vanishing longitudinal momentum
- the zero mode.

The canonical commutators of the field $\tilde A$ are \cite{burkardt}
\begin{equation}
[\tilde A^a_i(x^-,x),\tilde A^b_j(y^-,y)]=-\frac{i}{
2}\epsilon(x^--y^-)\delta^{ab}_{ij}(x-y)\, ,
\end{equation}
with
\begin{equation}
\epsilon(x)=\frac{1}{ 2}[\Theta(x)-\Theta(-x)]\, .
\end{equation}
One defines the light cone canonical creation and annihilation operators as usual through
\begin{eqnarray}
  &&\tilde A_i^a(x^-,x) = \int_0^\infty
  \frac{dk^+}{2\pi} \frac{1}{\sqrt{2k^+}}
  \Bigg\lbrace a_i^a(k^+, x)\, e^{-ik^+x^-}
  + a^a_{i\, \dagger}(k^+,x)\, e^{ik^+x^-}
  \Bigg\rbrace\, ,
  \nonumber \\
 &&\left[ a^a_i(k^+, x), a_j^{b\, \dagger}(p^+, y)
   \right] = (2\pi)\, \delta^{ab}\, \delta_{ij}\, \delta(k^+ -p^+)\,
             \delta^{(2)}(x - y)\, .
\end{eqnarray}
This translates into ($k^+\ne 0$):
\begin{equation}
  \left[ \tilde A^a_i(k^+,{ x}), \tilde A^b_j(p^+,{ y}) \right]=\frac{\pi}{ 2}\left(\frac{1}{k^++i\epsilon}+\frac{1}{k^+-i\epsilon}\right)\, \delta(k^+ +p^+)\, \delta^{ab}\, \delta_{ij}\,
             \delta^{(2)}({ x} - { y})\, .
%             \nonumber \\
%  \left[ \tilde A_a^i(k^+=0,{ x}), \partial^+\tilde A_b^j(p^+=0,{ y}) \right]
%  &=& i \frac{2}{\epsilon}\, \delta_{ab}\, \delta^{ij}\,
%             \delta^{(2)}({ x} - { y})\, .
\end{equation}
%
%recall that $\partial^+ \to -ik^+$.

The Hamiltonian eq.(\ref{hamiltonian}) commutes with the generator of the $x^-$ - independent gauge transformation, which on physical states should vanish:
\begin{equation}\label{elc}
\int dx^-({D}_i\partial^+A_i)^a-j^a(x)=0\, .
\end{equation}
Following the standard procedure we should fix this residual gauge freedom by imposing a gauge fixing condition.
We will be working in the gauge (same as in \cite{JIMWLK})
\begin{equation}
\partial_i\,A_i^a(x^-\rightarrow-\infty)=0\, .
\end{equation}

>From previous analysis of the behavior of the field in this gauge~\cite{JIMWLK},
 we know that the vector potential vanishes at $x^- \to - \infty$ but approaches a
 non vanishing asymptotic value at $x^-\rightarrow\infty$, which we denote by 
$A_i^a(x^-\rightarrow\infty,x)=\gamma_i^a(x)$. Separating the nonzero momentum modes,
we thus write
\begin{equation}
A_i^a(x^-,x)=\frac{1}{2}\gamma_i^a(x) +\tilde A^a_i(x^-,x)\, .
\end{equation}
Even though $\tilde A$ has no zero momentum mode, its asymptotics is
not vanishing but is rather given by $\pm\frac{1}{2}\gamma_i^a$. It is thus
convenient to define a field $c$ which has regular behavior at
infinity by
\begin{eqnarray}
&&A_i^a(x^-,x )\,=\,\theta(x^-)\gamma_i^a(x )
+c^a_i(x^-,x )\, , \nonumber \\
&&\tilde A_i^a(x^-,x )\,=\,\epsilon(x^-)\gamma_i^a(x ) +c^a_i(x^-,x)\, ,\nonumber \\
&&c^a_i(x^-\rightarrow\pm\infty,x) \rightarrow 0\, ,
\end{eqnarray}
and
\begin{equation}
\partial^+A_i^a=\partial^+\tilde A^a_i=\delta(x^-)\gamma_i^a+\partial^+c_i^a\, .
\end{equation}
Our aim is to find the ground state of the Hamiltonian
eq.(\ref{hamiltonian}) given the charge density $j^a$ (more precisely we consider
the matrix elements of the operators $j^a(x)$ on the  Hilbert space of the
valence modes as known). 

Our first observation is that since the Hamiltonian is the integral
of the positive definite Hamiltonian density over $x^-$, the
necessary condition for finiteness of energy is vanishing of the
density at $x^-\rightarrow\pm\infty$. The finiteness of the magnetic
part of the Hamiltonian requires
\begin{equation}\label{magnc}
G_{ij}^a(x^-\rightarrow\infty)=0\, ,
\end{equation}
while the finiteness of the electric part is ensured by eq.(\ref{elc}).

We will use the gauge fixing condition and the finite energy
conditions as operatorial constraints that determine $\gamma$ in
terms of $\tilde A$ (or equivalently $c_i$). This is equivalent to
Dirac bracket quantization of the fields $A$ which leave the 
canonical commutators of $\tilde A$ unchanged. The commutators of
$\gamma^a_i$ with $\tilde A^a_i$ and between themselves are then
determined by solving the  constraints.

Expressing the magnetic constraint equation (\ref{magnc}) in terms of $\gamma_i^a$, we obtain
\begin{equation}\label{constrmag}
\partial_i\,  \gamma^a_j(x) -
  \partial_j\, \gamma_i^a( x) - g f^{abc}\, \gamma^b_i( x)\,
                                    \gamma^c_j(x) = 0\, .
\end{equation}
To express the electric constraint eq.(\ref{elc}) we use
the fact that given the boundary conditions on $c_i^a$
\begin{equation}\label{intpi}
\int dx^-\partial^+A_i^a=\gamma_i^a\, .
\end{equation}
We then find\footnote{Here we used
$ f^{abc}\int dx^-A_i^b(x^-)\partial^+A^c_i(x^-)=\frac{1}{2}f^{abc}\gamma_i^b\gamma_i^c+f^{abc}\int dx^-\tilde
A_i^b(x^-)\partial^+\tilde A^c_i(x^-)$ which follows from
eq.(\ref{intpi}).}
\begin{equation}\label{constrel}
\partial_i\gamma_i^a-\frac{1}{2}gf^{abc}\gamma_i^b\gamma_i^c-gf^{abc}\int dx^-\tilde A_i^b(x^-)\partial^+\tilde A^c_i(x^-)=j^a(x)\, ,
\end{equation}
or, equivalently,
\begin{equation}\label{constrel1}
\partial_i\gamma_i^a(x)-\frac{1}{2}gf^{abc}\gamma_i^b(x)\gamma_i^c(x)+gf^{abc}\{\gamma_i^b(x),c^c_i(x,0)\}-gf^{abc}\int dx^-c_i^b(x^-)\partial^+c^c_i(x^-)=j^a(x)\, .
\end{equation}
In this equation $c(0)$ should be understood as
\begin{equation}
  c^a_i(x^-=0)\,=\,\tilde A(x^-=0)\,=\,\frac{1}{2}[c^a_i(x^-=0^+)+c^a_i(x^-=0^-)]\, ,
\end{equation}
where $0^+\equiv 0+\epsilon; \ \ 0^-\equiv 0-\epsilon; \ \ \epsilon\rightarrow 0$.
This is important since $c$ is not necessarily continuous at $x^-=0$.

%%%%%%%%%%%%%%%%%%%%%%%%%%%%%%%%%%%%%%%%%%%
\subsection{Solving the constraints}
\label{sec2.2}
 Our strategy now is the following. We should solve the two constraint equations,
eqs.(\ref{constrmag},\ref{constrel}) and determine the commutation
relations of the non dynamical field $\gamma$. Then we must
substitute it back into the Hamiltonian and express the Hamiltonian in
terms of the canonical degrees of freedom $\tilde A$.

We will do so by expanding the constraint equations and the Hamiltonian in powers of $g$. When doing so we must have some knowledge of the parametric dependence of the valence charge density $j$ on the coupling constant $g$. The expansion in principle can be performed for any parametric dependence.  In this section we take $j$ to be of order $1/g$ as in the JIMWLK limit, and will collect all contributions to the Hamiltonian of order $1/g$ and order $1$. 
It turns out however that this same resummation collects the leading terms in $g$ also for any 
$g\leq j \leq 1/g$. We will discuss this point in detail in the discussion section. Thus even though in this section we treat explicitly $j$ as being of order $1/g$ this should not be construed as limiting our calculation to the JIMWLK limit.

Thus our aim in this section is to expand $\gamma$ to $O(1)$, obtain the
Hamiltonian to $O(1)$ and diagonalize this $O(1)$ Hamiltonian
exactly. Further corrections to this calculation are strictly perturbative (small corrections in powers of
$g$ for any parametric dependence  of $j$ on $g$) and will not be considered
here.

To order $1/g$ the operator $\gamma$ satisfies the `classical equations' $\gamma^a_i=b^a_i$:
\begin{eqnarray}\label{classicaleq}
 && \partial_i\,  b^a_i( x) = j^a( x)\, ,
  \nonumber \\
 && \partial_i\,  b^a_j( x) -
  \partial_j\,  b^a_i( x) - g f^{abc}\, b_i^b( x)\,
                                    b_j^c( x) = 0\, .
\end{eqnarray}
To this order the commutation relations are calculated as
\begin{eqnarray}\label{commm}
  \left[ b^a_i( x), b^b_j( y) \right]
  &=& \int_{z,\bar z}\frac{\delta b^a_i( x)}{\delta j^c( z)}\,
  \left[ j^c(z), j^d(\bar z) \right]\,
   \frac{\delta b^b_j(y)}{\delta j^d(\bar z)}
   \nonumber\\
   &=& -i\,g  \,\int_{z}\left[ { D}_i\frac{1}{\partial   { D}}\right]^{ac}(x,z)\,
   f^{cde}\, j^e( z)\,
   \left[ \frac{1}{{ D}  \partial } { D}_j\right]^{db}( z, y)\, ,
\end{eqnarray}
where ${ D}$ is the transverse covariant derivative in the `classical' background field $b$: 
${ D}_i^{ab}=\partial_i\delta^{ab}-gf^{acb}b_i^c$.
Eq.(\ref{commm}) is the leading order result in $g$. Note however that it is exact in the weak field limit, where the field $b$ is linear in the valence charge density $j$.
 
Eq.(\ref{commm}) can be further simplified, using the identity
\begin{equation}
g\,f^{cde}\,j^e( z)\,=\,g\,f^{cde}\,\partial_ib_i^e(
z)\,=\,-\,[\partial_i,\partial_i-{ D}_i]^{cd}=(\partial { D}\,-\,{ D} \partial)^{cd}\, .
\end{equation}
Thus finally, to leading order in $g$
\begin{eqnarray}\label{comg}
&&  \left[ \gamma^a_i(x), \gamma^b_j( y)
  \right]=\left[ b^a_i( x), b^b_j( y) \right]
\,=\,-\,i\,d_{ij}^{ab}(x,y)\equiv i\,[{ D}_i\frac{1}{\partial   { D}}{ D}_j\,-\,
{ D}_i\frac{1}{{ D}  \partial
  }{ D}_j]^{ab}( x, y)\, ,
  \nonumber\\
&&[\gamma^a_i,\tilde A^b_j]\,=\,[b^a_i,\,\tilde A^b_j]\,=\,0\, .
\end{eqnarray}

Note that although $\gamma$ itself is of order $1/g$, the commutator
of two $\gamma$'s is of order one. It is thus clear that we will not
need higher order corrections  to the commutator eq.(\ref{comg}) in the $O(1)$ calculation.

To order $O(1)$ we write 
\begin{equation}\label{bdelta}
\gamma^a_i=b^a_i+\zeta^a_i\, ,
\end{equation}
where $\zeta$ satisfies the equations:
\begin{eqnarray}
&&\partial_i\zeta^a_i=-2gf^{abc}b_i^b\tilde A^c_i(x^-=0)\, ,
\nonumber\\
&&{ D}^{ab}_i\zeta^b_j-{ D}^{ab}_j\zeta^b_i=0\, .
\end{eqnarray}
The solution to these two equations is easily found as
\begin{equation}\label{delta}
\zeta_i^a=-2 \left[ { D}_i\frac{1}{\partial   { D}}(\partial-{ D})  \tilde A(x^-=0)\right]^a\, ,
\end{equation}
where the product on the right hand side is understood in the matrix
sense over all indexes (including transverse coordinates). Note that
the ordering of different factors of $b$ in eq.(\ref{delta}) is
irrelevant, since the covariant derivative involves $gb$, and the
commutator of two such factors is $O(g^2)$ and is thus of higher
order than the one we need to keep.

% For completeness we also give the expression for $\gamma$ to
%order $g$.
%\begin{equation}
%\gamma^a_i=b^a_i+\delta^a_i+\beta^a_i
%\end{equation}
%with
%\begin{eqnarray}
%&&\partial_i\beta^a_i+{i\over 4}gf^{abc}d^{bc}_{ii}(x,x)+gf^{abc}\{\delta^b_i,\tilde A^c_i(0)\}=gf^{abc}\int dx^-c_i^b\partial c^c_i\nonumber\\
%&&{ D}^{ab}_i\beta^b_j-{ D}^{ab}_j\beta^b_i=gf^{abc}\delta^b_i\delta^c_j
%\end{eqnarray}
%These equations are solved by
%\begin{eqnarray}\label{beta}
%&&\beta^a_i=-{g\over 2}\epsilon^{ij}\partial^j\left[{1\over { D}\partial}\right]^{ab}[f^{bcd}\epsilon_{kl}\delta^c_k\delta^d_l]
%+g\left[{ D}_i{1\over \partial { D}}\right]^{ab}\left[-{i\over 4}fd_{jj}(x,x)-f\{\delta_j,\tilde A_j(0)\}+f\int dx^-c_j^b\partial c^c_j\right]^b
%\end{eqnarray}
%When we get to calculating terms of $O(g)$ we should remember also to take $\tilde A$ to $O(g)$. That is the correlators of $\tilde A$ will have terms of $O(g)$ and those should be kept while calculating $\delta$ in eq.(\ref{delta}). These contributions however do not affect the expression for $\beta$ eq.(\ref{beta}), as it is already explicitly $O(g)$.

%For now we will forget about $O(g)$ terms, and will concentrate on the calculation at order $1$.

The canonical structure to $O(1)$ follows from
Eqs.(\ref{bdelta},\ref{delta})
\begin{eqnarray}\label{commutators}
&&[\gamma^a_i(x),\tilde A^b_j(y)]=[\zeta^a_i(x),\tilde A^b_j(y)]
=-i\epsilon(y^-)\left[{ D}_i\frac{1}{\partial { D}}(\partial -{ D})_j\right]^{ab}(x,y)\, ,\nonumber\\
&&[c_i^a(x),\gamma_j^b(y)]=i\epsilon(x^-)\left[\partial_i\frac{1}{{ D}\partial}{ D}_j
-{ D}_i\frac{1}{\partial { D}}{ D}_j\right]^{ab}(x,y)\, ,\nonumber\\
&&[c_i^a(x),c_j^b(y)]=-\frac{i}{2}\epsilon(x^--y^-)\delta^{ab}_{ij}(x-y)-\frac{i}{2}\epsilon(x^-)\epsilon(y^-) C^{ab}_{ij}(x,y)\, ,
%
% \left\{2{ D}_i{1\over
%\partial { D}}\partial_j-2\partial_i{1\over { D}\partial
%}{ D}_j\right\}^{ab}(x,y)
\end{eqnarray}
where for future convenience we have defined
\begin{equation}
C^{ab}_{ij}(x,y)=\left\{2\partial_i\frac{1}{{ D}\partial }{ D}_j-2{ D}_i\frac{1}{\partial { D}}\partial_j\right\}^{ab}(x,y)\, .
\label{2.29}
\end{equation}

%%%%%%%%%%%%%%%%%%%%%%%%%%%%%%%%%%%%%%%%%%%%%%%%%%%
\subsection{The Hamiltonian and the equations of motion to O(1)}
\label{sec2.3}
Next we express the Hamiltonian to O(1) in terms of the field
$c^a_i$.

For the magnetic piece to $O(1)$ we have:
\begin{equation}
G^{a}_{ij}(\theta(x^-)\gamma+c)
%{ D}_i\left(\theta(x^-)\gamma\right)c_j-{ D}_j\left(\theta(x^-)\gamma\right)                                                                    c_i-gf^{abc}c_i^bc_j^c
=\theta(-x^-)[\partial_ic_j-\partial_jc_i]+\theta(x^-)[{ D}_ic_j-{ D}_jc_i]+O(g)\, .
%-gf^{abc}c_i^bc_j^c+[gf^{abc}\delta^b_ic_j^c-gf^{abc}\delta^b_jc_i^c]\theta(x^-)\nonumber
\end{equation}
For the electric piece, using the constraint and after some algebra, we obtain:
\begin{equation}
\Pi^-_a=-\frac{1}{ \partial^+}[{ D}_i\partial^+A_i-j\delta(x^-)]^a=-\left[\theta(-x^-)\partial_ic_i+\theta(x^-){ D}_ic_i\right]^a+O(g)\, .
\end{equation}
All said and done the Hamiltonian to $O(1)$ is
\begin{equation}\label{hamiltonian1}
H\,=\,-\,\frac{1}{ 2}\,\int dx^-\,d^2x\,
\left[\theta(-x^-)\ c_i^a(x^-,x)\ \partial^2\ c_i^a(x^-,x)\ +\ \theta(x^-)\,c_i^a(x^-,x)\ { D}^{2\, {ab}}\ c_i^b(x^-,x)
\right]\, .
\end{equation}

This is the Hamiltonian that we have to diagonalize%\footnote{\it 
%Note that in the absence of the external field this Hamiltonian coincides with the free gluonic Hamiltonian.}
. 
The most efficient way of doing this is first to find the complete set of solutions of classical equations of motion, and then expand the quantum field operators in the canonical creation and annihilation operators with the coefficients given by the solutions of classical equations. The classical solutions have to be properly normalized in order that the quantum field operators  satisfy correct commutation relations. 

We start by deriving the equations of motion. Using the commutation relations eq.(\ref{commutators}) we obtain
\begin{eqnarray}\label{eqm}
i\partial^+\partial^-c^a_i(x)=[H,\partial^+c_i^a(x)]&=&\int dy^-[\partial^+c^a_i(x),c^b_j(y)]\left[\theta(-y^-)\partial^2+\theta(y^-){ D}^2\right]^{bc}_{jk}c^c_k(y)
\nonumber\\
&=&-\frac{i}{
2}[\theta(-x^-)\partial^2\delta^{ab}+\theta(x^-){ D}^{2\,ab}]c_i^b(x)
\nonumber \\
&&-\frac{i}{
4}\delta(x^-)C^{ab}_{ij}(x,y)\int
dy^-\left[-\theta(-y^-)\partial^2+\theta(y^-){ D}^2\right]^{bc}_{jk}c^c_k(y^-,y)\, , \nonumber \\
\end{eqnarray}
where $C^{ab}_{ij}$ is defined in (\ref{2.29}).
Integrating these equations (avoiding the singularity at $y^-=0$) gives
\begin{eqnarray}\label{int}
&&-\frac{i}{2}\int_{-\infty}^{0^-}dy^-\partial^2c(y)=i\int_{-\infty}^{0^-}dy^-\partial^+\partial^-c(y)=i\partial^-c(0^-)\, ,\\
\label{int1}
&&-\frac{i}{2}\int_{0^+}^\infty dy^-{ D}^2c(y)=i\int_{0^+}^\infty
dy^-\partial^+\partial^-c(y)=-i\partial^-c(0^+)\, .
\end{eqnarray}
The last term in eq.(\ref{eqm}) can be rewritten as
\begin{equation}\label{using}
-\frac{i}{4}\delta(x^-)C^{ab}_{ij}(x,y)\int dy^-\left[-\theta(-y^-)\partial^2+\theta(y^-){ D}^2\right]^{bc}_{jk}c^c_k(y)=
-i\delta(x^-)C^{ab}_{ij}(x,y)\partial^-c_j^b(0)\, ,
\end{equation}
so that finally the equations of motion are
\begin{equation}
i[\partial^++\delta(x^-)C]^{ab}_{ij}(x,y)\partial^-c_j^b(y)
=-\frac{i}{2}[\theta(-x^-)\partial^2+\theta(x^-){ D}^2]^{ab}(x,y)c_i^b(y)\, .
\end{equation}
Matching the discontinuity across $x^-=0$ gives the relation
\begin{equation}\label{discontinuity}
c_i^a(0^+,x)-c_i^a(0^-,x)=-\frac{1}{2}C^{ab}_{ij}(x,y)[c_i^b(0^+,y)+c_i^b(0^-,y)]\, .
\end{equation}
The solution to the equations of motion can be written down
explicitly. At negative $x^-$ this is just a free equation, and thus
the solution is a superposition of plane waves. At positive $x^-$
the solution is again a superposition of gauge rotated plane waves.
This can be written as
\begin{equation}\label{solu}
c_{i, p^-}^a(x)=\exp\{ip^-x^+\}\int
d^2q\left[\Theta(-x^-)\exp\{i\frac{\partial^2}{2p^-}x^-\}v^{i-}_{p^-q}(x)+\Theta(x^-)\exp\{i\frac{{ D}^2}{2p^-}x^-\}
v^{i+}_{p^-q}(x)\right]\, .
\end{equation}
Except at $x^-=0$ this solves the equations of motion with given $p^-$ for arbitrary $v^{i,\pm}_q$. Here $q$ is the degeneracy index. In the free theory the index $q$ would stand collectively for transverse momentum $k$, polarization index $i$ and color "polarization index" $a$. In the present case $q$ also stands for $i$ and $a$ as well as some continuous degeneracy. For simplicity of notation we will not differentiate between discrete and continuous parts of $q$. In the following, integral over $q$ stands both for the integral over continuous part with appropriate measure as well as for summation over the rotational and color 'polarizations'.

 Eq. (\ref{discontinuity}) imposes the condition
\begin{equation}\label{discontinuity1}
v_{i}^{a+}(x)-v_{i}^{a-}(x)=-\frac{1}{
2}C^{ab}_{ij}(x,y)[v_i^{b+}(y)+v_i^{b-}(y)]\, .
\end{equation}
This equation can be equivalently rewritten as
\begin{equation}\label{+-}
v^+_i=[T-L]^{ij}(t-l)^{jk}v^-_k\, ,
\end{equation}
where the projectors $T,\ L, t,\ l$ are defined as
\begin{equation}
L^{ab}_{ij}=\left[{ D}_i\frac{1}{ { D}^2}{ D}_j\right]^{ab},\ \ \ \ \ T^{ab}_{ij}=\delta^{ab}_{ij}-L^{ab}_{ij}; \ \ \ l_{ij}=\partial_i\frac{1}{\partial^2}\partial_j; \ \ \ \ t_{ij}=\delta_{ij}-l_{ij}\, .
\end{equation}
Eq.(\ref{+-}) is solved by
\begin{equation}
v^+_i=[T-L]^{ij}v_j; \ \ \ \ \ \ \ \  v^-_i=[t-l]^{ij}v_j
\end{equation}
for arbitrary $v_j$.
Thus we can write the solution eq.(\ref{solu}) in terms of one set
of functions $v^{ai}_{p^-q}(x)$ as
\begin{eqnarray}
c_{i, p^-}^a(x)&=&\exp\{ip^-x^+\}\int d^2q\left[\Theta(-x^-)\exp\{i\frac{\partial^2}{2p^-}x^-\}[t-l]_{ij}v^j_{p^-q}(x)
\right. \nonumber\\
&& \qquad \qquad \qquad \qquad  \left. +\Theta(x^-)\exp\{i\frac{{ D}^2}{2p^-}x^-\}[T-L]_{ij}v^j_{p^-q}(x)\right]\, .
\label{solu1}
\end{eqnarray}
On the level of the classical solution, the normalization of the
functions $v^{ai}_{p^-q}(x)$ is arbitrary. However, in order to use eq.(\ref{solu1}) as the basis for expansion of the operators $c$ in terms of
canonical creation and annihilation operators the
normalization of $v^{ai}_{p^-q}(x)$ has to be determined. This will be
done in the following subsection.

As a corollary to this subsection we note that the classical field $b$ does not commute
with the Hamiltonian and is therefore not constant in time.
Calculating the commutator we obtain
\begin{eqnarray}
&&i\partial^-b^a_i(x)\,=\,[H,b_i^a]\,=\,
\int dy^-[b^a_i(x),c^b_j(y)]\left[\theta(-y^-)\partial^2\,+\,\theta(y^-){ D}^2\right]^{bc}_{jk}\ c^c_k(y)
\nonumber\\
&&=\,
\frac{1}{2}\,d^{ba}_{ji}(y,x)\,
\int dy^-\left[-\theta(-y^-)\partial^2\,+\,\theta(y^-){ D}^2\right]^{bc}_{jk}\ c^c_k(y)\, .
\end{eqnarray}
Using eqs.(\ref{int},\ref{int1}) 
%\begin{equation}
%\int dy^-\left[\theta(-y^-)\partial^2+\theta(y^-){ D}^2\right]^{bc}_{jk}c^c_k(y)=4\partial^-c^b_j(0)
%\end{equation}
this can be written as
\begin{equation}
i\partial^-b^a_i(x)=2i\partial^-\left\{{ D}_i\frac{1}{{ D}\partial}{ D}_j-{ D}_i\frac{1}{\partial { D}}{ D}_j\right\}^{ab}c_j^b(0)\, .
\end{equation}
This can be interpreted in the following way. Let us define the operator $\bar b$, so that it has the same exact matrix
elements on the valence part of the Hilbert space as $b$, but
commutes with the operators $c$. Then we can write
\begin{equation}
b^a_i=\bar b^a_i+2\left\{{ D}_i\frac{1}{{ D}\partial}{ D}_j-{ D}_i\frac{1}{\partial { D}}{ D}_j\right\}^{ab}c_j^b(0)\, ,
\end{equation}
and
\begin{equation}\label{gammas}
\gamma^a_i=\bar b^a_i+2\left\{{ D}_i\frac{1}{ { D}\partial}{ D}_j-{ D}_i\frac{1}{\partial { D}}\partial_j\right\}^{ab}c_j^b(0)\, .
\end{equation}
This form will be convenient for calculating correlators of $\gamma$ in the vacuum state.
 
%%%%%%%%%%%%%%%%%%%%%%%%%%%%%%%%%%%%%%%%%%%%%%%%% 
\subsection{Normalization of the eigenfunctions and the vacuum state.}
\label{sec2.4}
Given that the $O(1)$ Hamiltonian is quadratic, and having found the
complete set of solutions of the classical equations of motion, we
can find the quantum vacuum state.

The vacuum state of the Hamiltonian eq.(\ref{hamiltonian1}) is the
Fock vacuum of the canonical annihilation operators
$\beta_{p^-,q}$ defined in terms of $c$ by
%\footnote{A word of caution: this $\alpha$ has nothing to do with  the target color field $\alpha_T$ introduced in eq. (1.2).}
%
\begin{eqnarray}
c_{i}^a(x)&=&\int_{0}^\infty \frac{dp^-}{ 2\pi}\int d^2q \left[\Theta(-x^-)e^{i\frac{\partial^2}{2p^-}x^-}[t-l]_{ij}(x,y)v^{aj}_{p^-,q}(y)\right.
\nonumber \\
&& \left. \qquad \qquad \qquad \qquad +\Theta(x^-)e^{i\frac{{ D}^2}{2p^-}x^-}[T-L]^{ab}_{ij}(x,y)v^{bj}_{p^-,q}(y)\right]\beta_{p^-,q}+h.c. \, ,
\label{canonical}
\end{eqnarray}
where the integral over the transverse coordinate $y$ is understood but not written explicitly.
The operators $\beta$ satisfy canonical commutation relations
\begin{equation}
[\beta_{p^-,q},\,\beta^{\dagger}_{p'^-,q'}]\,=\,(2\,\pi)\,\delta(p^--p'^-)\,\delta(q-q')\, .
\end{equation}
Existence of such a set of canonical operators is guaranteed if the
set of solutions of the classical equation is complete and the
functions $v$ entering eq.(\ref{canonical}) are properly normalized.
To find the correct normalization of these functions we require that
$c$ satisfy eq.(\ref{commutators}).

We concentrate on negative $x^-$ and $y^-$ first, so that only the first term in the sum in eq.(\ref{canonical}) is important. For simplicity we suppress the color indexes and also the factor $t-l$, thus we are working in terms of $v^-$ rather than $v$. Consider the commutator
\begin{eqnarray}\label{ccom}
[c^i(x),c^j(y)]&=&\int_{0}^\infty \frac{dp^-}{ 2\pi}\left[e^{i\{\frac{\partial_x^2}{
2p^-}x^--\frac{\partial_y^2}{
2p^-}y^-\}}\int_q v_{p^-,q}^{-i}(x)v_{p^-,q}^{*-j}(y)-e^{-i\{\frac{\partial_x^2}{
2p^-}x^--\frac{\partial_y^2}{
2p^-}y^-\}}\int_q v_{p^-,q}^{*-i}(x)v_{p^-,q}^{-j}(y)\right]\nonumber\\
&=&\int_{0}^\infty \frac{dp^-}{ 2\pi}\left[e^{i\{\frac{\partial_x^2}{2p^-}x^--\frac{\partial_y^2}{
2p^-}y^-\}}{\cal W}^{ij}_{p^-}(x,y)-e^{-i\{\frac{\partial_x^2}{2p^-}x^--\frac{\partial_y^2}{2p^-}y^-\}}
{\cal W}^{*ij}_{p^-}(x,y)\right]\, .
\end{eqnarray}
We have defined the `correlator matrix'
\begin{equation}
{\cal W}^{ij}_{p^-}(x,y)\,=\,\int {d^2q}\,v_{p^-,q}^{-i}(x)\,v_{p^-,q}^{*-j}(y)\, .
\end{equation}
Note that this matrix fully determines the commutators of $c$,
and there is no need to find the individual functions $v_{p^-,q}$.
Different choices of the functions $v$ which give the same $\cal W$
correspond to unitary rotations of the set of the canonical
operators $\beta$.

To determine the correct normalization we first note that taking 
\begin{equation}
{\cal W}^{ij}_{p^-}(x,y)\,=\,\delta^{ij}\,\delta^2(x-y)\,\frac{1}{2\, p^-}
\end{equation}
would give canonical commutation relations for the fields $c$. With this
expression for $\cal W$ we can change variables $p^-\rightarrow -p^-$ in the
second term of eq.(\ref{ccom}) to get
\begin{equation}
[c^i(x),c^j(y)]=\int_{-\infty}^\infty \frac{dp^-}{ 4\,\pi\, p^-}e^{i\{\frac{\partial^2}{
2p^-}x^--\frac{\partial^2}{
2p^-}y^-\}}\delta^2(x-y)\delta^{ij}=-\frac{i}{2}\delta^{ij}\delta^2(x-y)\epsilon(x^--y^-)\, ,
\end{equation}
where the last line follows by change of variables $p^-\rightarrow \partial^2/2p^-$.
To get the $\epsilon$-function in the commutator we have to regulate the singularity in  $1/p^-$ in the symmetric way
\begin{equation}
\frac{1}{ p^-}\rightarrow \left(\frac{1}{p^-}\right)^2\left[\frac{1}{\frac{1}{p^-}+i\epsilon}+\frac{1}{\frac{1}{p^-}-i\epsilon}\right]\, .
\end{equation}
To reproduce the extra term in the commutator of $c^i$ (the second term in
the last line of eq.(\ref{commutators})) we modify the matrix $\cal W$ in
the following way
\begin{equation}\label{M}
{\cal W}^{ij}_{p^-}(x,y)={1\over 2}\,
\left(\frac{1}{p^-}\right)^2\left\{\frac{1}{\frac{1}{p^-}+i\epsilon}[\delta^{ij}\delta^2(x-y)+\frac{1}{2}C^{ij}(x,y)]+\frac{1}{\frac{1}{p^-}-i\epsilon}[\delta^{ij}\delta^2(x-y)-\frac{1}{2}C^{ij}(x,y)]\right\}\, .
\end{equation}
The new term we have added is imaginary and even with respect to
$p^-\rightarrow -p^-$. Thus it is still true that the two terms in
eq.(\ref{ccom}) are equal. The extra term under the change of
variables $p^-\rightarrow 1/p^-$ gives
\begin{equation}
\int d\left(\frac{1}{ p^-}\right) \delta\left(\frac{1}{p^-}\right)\, ,
\end{equation} 
and thus generates the term in the commutator independent of $x^-$ and $y^-$.
The result is precisely the last term of eq.(\ref{commutators}). It is a
matter of some straightforward algebra to check that with ${\cal W}$
defined in eq.(\ref{M}) the correct commutator of the fields $c$ is
reproduced also for other values of $x^-$ and $y^-$. The following
identities come handy in this calculation
\begin{eqnarray}\label{identity}
&&1-\frac{1}{ 2}C=\left[1+\frac{1}{ 2}C\right](T-L)(t-l);\ \ \ \ \ \left[1-\frac{1}{ 2}C\right](t-l)=\left[1+\frac{1}{ 2}C\right](T-L)\, ;\nonumber\\
 %&&1+{1\over 2}C=\left[1-{1\over 2}C\right](t-l)(T-L); \ \ \ \ \left[1+{1\over 2}C\right](T-L)=\left[1-{1\over 2}C\right](t-l)\, ;\nonumber\\
 &&(t-l)C(t-l)=-C; \ \ \ \ \ (T-L)C(T-L)=-C\, .
\end{eqnarray}

Returning from $v^-$ to $v$ we conclude that the operators $\beta$,
$\beta^\dagger$ in the representation eq.(\ref{canonical}) have
canonical commutation relations when (we use eq.(\ref{identity}))
\begin{eqnarray}
\int d^2q\,v_{p^-q}^{i}(x)\,v_{p^-q}^{*j}(y)&=&{1\over 2}\,
\left(\frac{1}{ p^-}\right)^2\left\{\left[\frac{1}{\frac{1}{p^-}+i\epsilon}
+\frac{1}{ \frac{1}{ p^-}-i\epsilon} \right]\delta^{ij}\delta^2(x-y) \right.
\nonumber \\
&&  \left. \qquad \qquad \quad 
-\frac{1}{ 2}\left[\frac{1}{\frac{1}{ p^-}+i\epsilon}-\frac{1}{\frac{1}{ p^-}-i\epsilon} \right]C^{ij}(x,y)\right\}\, .
\label{vnorm}
\end{eqnarray}

We thus conclude that the vacuum of the Hamiltonian
eq.(\ref{hamiltonian}) to $O(1)$ is the Fock vacuum of the
annihilation operators $\beta$ related to the original gluon field
operators through
\begin{eqnarray}\label{ac}
\tilde A^a_i(x^-,x)&=&\epsilon(x^-)\left[b^a_i(x)-2{ D}_i\frac{1}{
\partial
{ D}}(\partial-{ D})(x,y)c(0,y)\right]+c^a_i(x^-,x)\nonumber\\
&=&\epsilon(x^-)\left[\bar b^a_i(x) +2\left\{{ D}_i\frac{1}{{ D}\partial}{ D}_j-{ D}_i\frac{1}{\partial
{ D}}\partial_j\right\}^{ab}(x,y)c_j^b(0,y) \right]+c^a_i(x^-,x)
\end{eqnarray}
with the field $c^a_i(x^-,x)$ expressed in term of $\beta$ and
$\beta^\dagger$ in eq.(\ref{canonical}) with the normalization
eq.(\ref{vnorm}).

This completes the diagonalization of the light cone Hamiltonian to $O(1)$.
%We observe that the contribution that is responsible for the
%additional term in the commutator comes from the mode with infinite
%energy
%\begin{equation}
%\left[\theta(-x^-)(t-l)+\theta(x^-)(T-L)\right] v_\infty
%\end{equation}
%If we omit the contribution of this mode however, the commutators of $c$ still do not become canonical.

%%%%%%%%%%%%%%%%%%%%%%%%%%%%%%%%%%%%%%%%%%%%%%%%%%%
\section{The Bogoliubov operator}
The calculation of the previous section can be viewed as the diagonalization of the light cone 
Hamiltonian. Although we have only found the vacuum state, quite generally the diagonalization 
is affected by the action of some unitary operator $\Omega$. Namely for the case of a quadratic 
operator $H$
\begin {equation}
\Omega^\dagger H\Omega=\int
_{p^-,q}p^-\beta^\dagger_{p^-,q}\beta_{p^-,q}\, .
\end{equation}
The explicit knowledge of the operator $\Omega$, or alternatively the knowledge of its action on all 
the degrees of freedom of the theory furnishes much more information than just the vacuum wave
function, as it also in principle can give us the wave functions of excited states, which are necessary 
to calculate more exclusive properties than the forward scattering amplitude. The aim of this section is 
to find explicitly the action of $\Omega$ on the degrees of freedom of the theory.

Part of the answer to this question is already furnished by
eq.(\ref{ac}) which can be viewed as the transformation of the
vector potential if we read the left hand side as $\Omega^\dagger \tilde A\Omega$ 
and the canonical operators $\beta$ and $\beta^\dagger$ in $c$ on the right hand side as the original gluon creation and annihilation operators $a$ and $a^\dagger$. The missing piece of information is the
transformation of the valence charge density. This is the question we address now.

First, it is clear from eq.(\ref{ac}) that the transformation is of the Bogoliubov form, namely 
\begin{equation}
\Omega\equiv {\cal C}\, {\cal B} = \exp\left[E\,\tilde A\right]\ 
\exp\left[{1\over 2}\,\tilde A\,M\,\tilde A\right]\, ,
\end{equation}
 where $E$ and $M$ are operators which depend on the charge density $j$ but do not depend on the soft fields $A$. We do not indicate explicitly the indexes and coordinate dependences of $E$ and $M$ for simplicity. Those should be clear from the context. Here $\cal C$ is a purely coherent state operator - exponent of an operator linear in $\tilde A$, while ${\cal B}$ has no linear term in the exponent.
The coherent operator is easy to find by inspection, since it is the only one that induces the shift of the soft field (the very first term in eq.(\ref{ac})):
\begin{equation}\label{coher}
{\cal C}=\exp\left[ 2\,i\,\int d^2x\,b^a_i(x)\,\tilde A^a_i(x^-=0,x)\right]\, .
\end{equation}
The Bogoliubov part of the transformation, the operator ${\cal B}$ is more difficult to determine. Rather than looking for the explicit form of the operator ${\cal B}$ in terms of $j$, we will find its action on the degrees of freedom of the theory by considering sequential action of ${\cal C}$ and ${\cal B}$ on $\tilde A$ 
and matching it onto eq.(\ref{ac}). 

It is important to remember that we need to know the transformation of the color charge density only to $O(g)$. Only this order contributes to the JIMWLK evolution as explained in detail in \cite{JIMWLK}. Thus we will determine the action of ${\cal B}$ on the fields to this order only.

We first note the following 'combinatorial' identity. For any operators $O$ and $L$
\begin{equation}\label{trans}
e^{-L}Oe^L=O+[O,L]+\frac{1}{2}[[O,L],L]+\frac{1}{3!}[[[O,L],L],L]+...
\end{equation}

Using eq.(\ref{trans}), we have for $\cal C$ of eq.(\ref{coher})
\begin{eqnarray}
{\cal C}^\dagger\, \tilde A_i^a(x)\,{\cal C} &=&
\tilde A_i^a(x)\ +\ \epsilon(x^-)\,b^a_i(x)\ +\ \epsilon(x^-)\,\int_y d^{ab}_{ij}(x,y)\,\tilde A^b_j(y^-=0,y)
\nonumber \\
&& +\,\frac{2\,i}{ 3}\,\epsilon(x^-)\,
\int_{y,z}[d^{ab}_{ij}(x,y),\,b^c_k(z)]\ \tilde A^b_j(y^-=0,y)\ \tilde A^c_k(z^-=0,z)\, ,
\label{atrans}\\
{\cal C}^\dagger \,j^a(x)\,{\cal C}&=&j^a(x)\ +\ 2\, \int_y 
\left\{\left(\partial { D}\frac{1}{ { D}\partial}-1\right){ D}_j\right\}^{ab}(x,y)\ 
\tilde A^b_j(y^-=0,y)\nonumber \\
&&
+\,2\,i\int_{y,z} \left[\left\{\left(\partial { D}\frac{1}{ { D}\partial}-1\right){ D}_j\right\}^{ab}(x,y),b^c_k(z)\right]
\tilde A^b_j(y^-=0,y)\tilde A^c_k(z^-=0,z)\, .
\label{jtrans} 
%C^\dagger b_i^a(x)C=b^a_i(x)+2d^{ab}_{ij}(x,y)\tilde A^b_j(y,y^-=0)+2i[d^{ab}_{ij}(x,y),b^c_k(z)]\tilde A^b_j(y,y^-=0)\tilde A^c_k(z,z^-=0)
\end{eqnarray}

To find the action of the Bogoliubov operator, we imagine diagonalizing the Hamiltonian first by acting with ${\cal C}$ and then subsequently acting with ${\cal B}$. Transforming the Hamiltonian eq.(\ref{hamiltonian}) with ${\cal C}$ obviously leads to
\begin{equation}\label{triv}
{\cal C}^\dagger \ H[\tilde A,\,j]\ {\cal C}\ \equiv \ H'[\tilde A,\,j]\ =\ 
H[{\cal C}^\dagger\, \tilde A \,{\cal C},\,{\cal C}^\dagger \,j\,{\cal C}]\, .
\end{equation}
It is straightforward to see using the expression for the transformed fields eqs.(\ref{atrans},\ref{jtrans}), that if we substitute for $\tilde A$ in the function $H'$ the following expression
\begin{equation}
\tilde A^a_i(x)\rightarrow c^a_i(x)\ +\ \epsilon(x^-)\,\Delta^{ab}_{ij}(x,y)\ c^b_j(y^-=0,y)\, ,
\end{equation}
with
\begin{equation}
\Delta^{ab}_{ij}(x,y)=\left\{{ D}_i\frac{1}{ \partial { D}}{ D}_j+{ D}_i\frac{1}{{ D}\partial }{ D}_j-2{ D}_i\frac{1}{\partial { D}}\partial_j\right\}^{ab}(x,y)\, ,
\end{equation}
we obtain to $O(1)$ precisely eq.(\ref{hamiltonian1}).
This substitution should be equivalent to the action of the Bogoliubov operator
\begin{equation}\label{bh}
{\cal B}^\dagger \,H'[\tilde A,\,j]\,{\cal B}\ =\ 
H'[{\cal B}^\dagger\, \tilde A\, {\cal B},\,{\cal B}^\dagger\, j\,{\cal B}]
\ \equiv\ 
H''[\tilde A,\, j]\, .
\end{equation}
In other words,
 up to (and including) $O(g)$ terms  the action of the Bogoliubov operator $\cal B$ on the field $\tilde
A$ is
\begin{equation}\label{transA}
{\cal A}^a_i(x,j)\ \equiv\  {\cal B}^\dagger\, \tilde A^a_i(x)\,{\cal B}\ =\ 
c^a_i(x)\ +\ \epsilon(x^-)\,\Delta^{ab}_{ij}(x,y)\ c^b_j(y^-=0,y)\, ,
\end{equation}
where the field $c$ on the RHS is understood as expressed in terms of the canonical creation and annihilation operators $a$ and $a^\dagger$ (rather than $\beta$ and $\beta^\dagger$)\footnote{We note that strictly speaking to make this identification we should also substitute into $H'$ the transformed expression for $j$ in eq.(\ref{bh}), which we do not know at this point. However as we will see below and is simple to understand by straightforward counting of powers of $g$, the operator $B$ induces transformation of $j$ only to order $g$. Since we only need the Hamiltonian to $O(1)$ it is therefore perfectly consistent to keep $j$ unchanged in $H'$ eq.(\ref{bh}) for the purpose of the identification of the Bogoliubov transformation of $\tilde A$.}. 

%This expression is of course consistent with what we expect. Our expectation is that the Bogoliubov operator has the general form
%\begin{equation}
%B=\exp\{iM(D)\tilde A^2\}
%\end{equation}
%Thus it should transform $\tilde A$ as
%\begin{equation}\label{bog}
%B^\dagger \tilde AB=Z(D)\tilde A
%\end{equation}
%where $Z$ is some functional of the background covariant derivative $D$. Corrections to this relation come from commutators between the covariant derivatives, which are $O(g^2)$. Sice we are only interested in $O(g)$ this is all we need. The only piece of information that we are missing therefore is how the Bogoliubov operator tranforms the backgroud field $b$. Here eq.(\ref{bog}) suggests the tranformation of the form
%\begin{equation}
%B^\dagger bB=b+gLA^2+O(g^2)
%\end{equation}

Our aim is now to find the transformation of the color charge density $j^a$ under the Bogoliubov transformation which induces eq.(\ref{transA}).
This is indeed possible, even though we do not know the explicit form of the operator $\cal B$
 itself in terms of the fundamental fields. The key is given by the following chain of arguments.

Consider a general Bogoliubov operator of the form
\begin{equation}\label{gob}
{\cal B}=\exp\left[ \frac{1}{2}\tilde A_i\,M_{ij}\,\tilde A_j\right]\, .
\end{equation}
Here we denote all indexes/coordinates of the field $A$ by a single index $i$. 
The fields $A$ are assumed to satisfy the commutation relation
\begin{equation}
[\tilde A_i,\,\tilde A_j]\ =\ P_{ij}
\end{equation}
with some matrix $P$. Quite generally the matrix $M$ is symmetric and anti hermitian, while $P$ 
is antisymmetric. The matrix $M$ depends on the charge density and the coupling constant only 
through the combination $g\,j$.

Consider the transformation
\begin{equation}\label{bara}
{\cal A}_k\equiv {\cal B^\dagger}\, \tilde A_k\,{\cal B}\,
=\,\tilde A_k+(PM\tilde A)_k+\frac{1}{ 2}(PMPM\tilde A)_k+\frac{1}{3!}(PMPMPM\tilde A)_k+...
=\,[e^{PM}]_{kl}\,\tilde A_l\, .
\end{equation}
Here we have used the identity eq.(\ref{trans}). Also, consistently with our counting of powers of the coupling constant we have neglected all and any terms involving commutators of $gj$ which enter into $M$, since each such commutator brings a power $g^2$.

Now to order $O(g)$ we have
\begin{equation}
[j^a,M_{ij}]=igf^{abc}j^c\frac{\partial M_{ij}}{\delta j^b}\, .
\end{equation}
Thus consider the transformation of $j^a(x)$ induced by the action of $\cal B$ in eq.(\ref{gob}):
\begin{eqnarray}\label{transjgen}
{\cal B}^\dagger j^a{\cal B}=j^a&+&\frac{i}{2}gf^{abc}j^c\Bigg\{\tilde A\frac{\delta M}{\delta j^b}\tilde A+
\frac{1}{2}\tilde A\left(\frac{\delta M}{\delta j^b}PM-MP\frac{\delta M}{\delta j^b}\right)\tilde A\\
&+&\frac{1}{ 3!}\tilde A\left(\frac{\delta M}{ \delta j^b}PMPM+
MPMP\frac{\delta M}{ \delta j^b}-2MP\frac{\delta M}{\delta j^b}PM\right)\tilde A+...\Bigg\}\, .
\nonumber
\end{eqnarray}
Here again we neglected all commutators of $gj$ in $M$ beyond the first term, as they are all higher order in $g$.
The negative signs come from transposing the antisymmetric matrix $P$.
We can now check explicitly that eq.(\ref{transjgen}) is expansion in powers of $M$ of the following expression
\begin{equation}
j^a+\frac{i}{2}gf^{abc}j^c{\cal A}P^{-1}\frac{\delta {\cal A}}{\delta j^b}
\end{equation}
with $\cal A$ defined in eq.(\ref{bara}).
Remembering that in our case $P=\frac{i}{ 2}\epsilon(x^--y^-)$ whose inverse is $-2i\partial^+$, and restoring all the indexes and coordinate dependences we obtain
\begin{equation}\label{transj}
\bar j^a(x)\equiv {\cal B}^\dagger j^a(x){\cal B}=j^a(x)+gf^{acd}j^d(x)\int dy^-d^2y\,
\partial^+{\cal A}^b_j(y^-,y)\frac{\delta {\cal A}^b_j(y^-,y)}{\delta j^c(x)}\, 
\end{equation}
with $\cal A$ given in eq.(\ref{transA}).

An equivalent way of obtaining this result is to require that the transformed fields satisfy the same commutation relations as the non transformed ones, the transformation being unitary. 
Using the explicitly known commutator of the field $c$ one can easily show that
\begin{eqnarray}
&&[{\cal A}^a_i(x^-,x),{\cal A}^b_j(y^-,y)]=-\frac{i}{ 2}\epsilon(x^--y^-)\delta^{ab}_{ij}(x-y)\, ,\\
&&{[j^a(x),{\cal A}^b_j(y^-,y)]} =\int_z[j^a(x),j^c(z)]\frac{\delta {\cal A}^b_j(y^-,y)}{\delta j^c(z)}=igf^{acd}j^d(x)\frac{\delta {\cal A}^b_j(y^-,y)}{\delta j^c(x)}\, .\nonumber
\end{eqnarray}
In this expression we should understand $\cal A$ as a function of $j$ at fixed $a$.
It is easy to check that with the transformation eq.(\ref{transj}) to $O(g)$
\begin{equation}
[\bar j^a(x),{\cal A}^b_j(y^-,y)]=0\, .
\end{equation}
This is straightforward after noticing that the last term in eq.(\ref{transj}) can be written as
\begin{equation}
%\bar j^a(x)&=&j^a(x)-i\int dy^-dydz\partial^+
g\,f^{acd}\,j^d(x)\int dy^-d^2y\partial^+{\cal A}^b_j(y^-,y)\frac{\delta {\cal A}^b_j(y^-,y)}{\delta j^c(x)}=
\int dy^-d^2y\,d^2z\,{\cal A}^b_j(y^-,y)[j^a(x),j^c(z)]\frac{\delta {\cal A}^b_j(y^-,y)}{ \delta j^c(z)}\, .
%&=&j^a(x)+gf^{acd}j^d(x)\int dy^-dy\partial^+\bar A^b_j(y^-,y){\delta \bar A^b_j(y^-,y)\over \delta j^c(x)}\, ,
\end{equation}
Therefore we conclude that the transformation eqs.(\ref{transA},\ref{transj}) does indeed preserve canonical commutation relations of the fields.

We can now put all the elements together and write down the transformation that the operator $\Omega$ induces on the fields:
\begin{eqnarray}
\Omega^\dagger \tilde A^a_i(x^-,x)\Omega 
&=& c^a_i(x^-,x)+\epsilon(x^-)\left[ b^a_i(x) +2\int_y\left\{{ D}_i\frac{1}{{ D}\partial}{ D}_j
                                       -{ D}_i\frac{1}{\partial { D}}\partial_j\right\}^{ab}(x,y)c_j^b(0,y) \right]
\nonumber\\
&&+\epsilon(x^-)\int_{y,z}\left\{g\left[{ D}_i\frac{1}{ \partial { D}}\right]^{ab}(xz)f^{bcd}j^d(z)
\int dy^-\partial^+{\cal A}^e_j(y^-,y)\frac{\delta {\cal A}^e_j(y^-,y)}{\delta j^c(z)}
\right. 
\nonumber \\
&& \left. +\frac{2i}{3}[d^{ab}_{ij}(x,y),b^c_k(z)]c^b_j(0,y)c^c_k(0,z)\right\}\, ,
\label{atrans1}\\
\Omega^\dagger j^a(x)\Omega &=& 
j^a(x)+2\int_y\left\{\left(\partial { D}\frac{1}{ { D}\partial}-1\right){ D}_j\right\}^{ab}(x,y) c^b_j(0,y)
\nonumber \\
&&+gf^{acd}j^d(x)\int dy^-d^2y\partial^+{\cal A}^b_j(y^-,y)\frac{\delta {\cal A}^b_j(y^-,y)}{ \delta j^c(x)}
\nonumber \\
&&+2i\int_{y,z}\left[\left\{\left(\partial { D}\frac{1}{ { D}\partial}-1\right){ D}_j\right\}^{ab}(x,y),b^c_k(z)\right] 
c^b_j(0,y)c^c_k(0,z)\, .\label{jtrans1}
\end{eqnarray}
Here $\cal A$ is given by eq.(\ref{transA}) and the field $c$ is understood as expressed in terms the canonical creation and annihilation operators $\beta$ and $\beta^\dagger$ as in eq.(\ref{canonical}). The first line of eq.(\ref{atrans1}) coincides with eq.(\ref{ac}). The second and third lines are the $O(g)$ terms. They are given here for completeness even though they do not contribute in the calculation of the previous section and also do not contribute to the transformation of the soft color charge density eq.(\ref{J}). 

Eqs.(\ref{atrans1},\ref{jtrans1}) are the main result of this section. They give the explicit action of the diagonalizing operator $\Omega$ on the fundamental degrees of freedom of the theory.

Finally, for completeness we give the expression for the transformation of the total charge density.
This is the observable directly relevant for the calculation of the scattering amplitude. It includes the contribution of the valence and the soft modes
\begin{equation}
J^a(x)=j^a(x)+gf^{abc}\int dx^-\tilde A^b_i(x)\partial^+\tilde A^c_i(x)\, .
\end{equation}
Collecting the formulae given above we find
%To find its correlators in the vacuum state we have to transform it with the coherent and the Bogoliubov transformations in turn
%\begin{eqnarray}
%&&C^\dagger J^a(x)C=j^a(x)+2g\epsilon^{acd}j^d(x)e^{bc}_j(yx)\tilde A^b_j(y,0)+2gi\left[\epsilon^{aed}j^d(x)e^{be}_j(yx),b^c_k(z)\right]\tilde A^b_j(y,0)\tilde A^c_k(z,0)\nonumber\\
%&&+gf^{abc}\int dx^-\tilde A^b_i(x)\partial^+\tilde A^c_i(x)+gf^{abc}b_i^b(x)\left[\int dx^-\epsilon(x^-)\partial^+\tilde A_i^c(x)-\tilde A_i^c(x)\right]
%\end{eqnarray}
%Finally we perform the Bogoliubov transform and find
% This amounts to the transform $\tilde A(0)\rightarrow c(0)$. Also the $\epsilon$ piece of the relation  between $\tilde A$ and $c$ is not relevant in the last term. Finally
%\begin{equation}
%\int dx^-\epsilon(x^-) \partial^+\tilde A_i^c(x)\rightarrow \int dx^-\epsilon(x^-) \partial^+c_i^c(x)=-c_i^c(x,0)
%\end{equation}
%Thus
\begin{equation}
\Omega^\dagger J^a(x)\Omega = j^a(x)+\delta_1j^a(x)+\delta_2j^a(x)\, ,
\label{J}
\end{equation}
with
\begin{eqnarray}
\delta_1j^a(x) &=& 2\left[\partial { D}\frac{1}{{ D}\partial}{ D}_j-\partial_j\right]^{ab}(x,y)c_j^b(0,y)\, ,
\label{j1}\\
\delta_2j^a(x)&=&2gi\left[f^{aed}j^d(x)e^{be}_j(y,x),b^c_k(z)\right]c^b_j(0,y)c^c_k(0,z)
\nonumber \\
&& +gf^{abc}\int dx^-{\cal A}^b_i(x)\partial^+{\cal A}^c_i(x)\nonumber\\
&& + gf^{acd}j^d(x)\int dy^-\partial^+{\cal A}^b_j(y^-,y)\frac{\delta {\cal A}^b_j(y^-,y)}{ \delta j^c(x)}\, ,
\label{j2}
\end{eqnarray}
with $\cal A$ given by eq.(\ref{transA}). Here
\begin{equation}
e_i^{ab}(x,y)=\frac{\delta b^a_i(x)}{ \delta j^b(y)}=\left[{ D}_i\frac{1}{ \partial { D}}\right]^{ab}(x,y)\, .
\end{equation}
As a consistency check with the calculation of the previous section we note that eq.(\ref{j1}) coincides with the divergence of eq.(\ref{gammas}).

%For reference in Appendix A we give the expression in terms of $c$ whose corelation functions are explicitly calculable.

%%%%%%%%%%%%%%%%%%%%%%%%%%%%%%%%%%%%%%%%%%%%%%%%%%
\section{Reproducing JIMWLK/KLWMIJ.}
As a cross check on our derivation we reproduce in this section the two known limits of the high energy evolution -  the JIMWLK evolution equation (the high density limit) and the KLWMIJ evolution equation (the low density limit)
. 
\subsection{The JIMWLK kernel}
Under boost the color charge density $j$ transforms into $J$ of eq.(\ref{J}). To derive the evolution of the functional $W$ we have to calculate the correlation functions of $J$ over the soft gluon vacuum, that is over the Fock vacuum of operators $\beta$. In the JIMWLK limit it is only necessary to know two correlators,
\begin{equation}\label{chi0}
\chi^{ab}(x,y)\equiv\lim_{\Delta Y\rightarrow 0}
{\langle 0_\beta|\,\delta_1j^a(x)\delta_1j^b(y)\,|0_\beta\rangle\over \Delta Y},\ \ \ \ \ \ \ \ \sigma^a(x)=\lim_{\Delta y\rightarrow 0}{\langle 0_\beta|\, \delta j_2^a(x)\,|0_\beta\rangle\over \Delta Y}\ ,
\end{equation}
 since $\delta_1j\sim gj$ and $\delta_2j\sim g^2j$, and so only these two correlators contribute to the evolution of $\langle j(x_1)...j(x_n)\rangle$ to relative order $g^2$. In fact our task is somewhat easier, since we can avoid the calculation of $\langle \delta j_2^a(x)\rangle$ using the following argument. In terms of $\chi$ and $\sigma$ the evolution kernel has the form
\begin{equation}
H^{JIMWLK}=\frac{1}{ 2}\chi^{ab}(x,y)\frac{\delta}{\delta j^a(x)}\frac{\delta}{\delta j^b(y)}
+\sigma^a(x)\frac{\delta}{\delta j^a(x)}\, .
\end{equation}
However it was proved in \cite{yinyang} that the evolution kernel has to be a Hermitian
operator (on the space of functions of $j$). In conjunction with the fact that $\sigma^a(x)$ is real, since it is a diagonal matrix element of an Hermitian operator (on the QCD Hilbert space), it means that $\sigma$ is rigidly related to $\chi$ so that the evolution kernel is
\begin{equation}\label{herm}
H^{JIMWLK}=\frac{1}{2}\frac{\delta}{\delta j^a(x)}\chi^{ab}(x,y)\frac{\delta}{\delta j^b(y)}\, .
\end{equation}
This property of the JIMWLK kernel is of course well known and has been first noted by Weigert in the last reference in \cite{JIMWLK}. Thus our task is first to calculate $\chi^{ab}(x,y)$ and then to show that the resulting evolution equation is equivalent to the standard form of JIMWLK which involves derivatives with respect to the unitary matrices $U$ rather than with respect to the charge density $j$.

We start with the calculation of $\chi$, defined as eq.(\ref{chi0}).
In preparation we calculate
\begin{eqnarray}
{1\over \Delta Y}\ \langle 0_\beta|\, c^a_i(0,x)c^b_j(0,y)\,|0_\beta \rangle &=& 
{1\over 8\,\Delta Y}\,\int \frac{dp^-}{ 2\pi p^-}[t-l+T-L][t-l+T-L]^{ab}_{ij}(x,y) 
\nonumber \\
&=& \frac{1}{ 4\,\pi}\ [1\,-\,l\,-\,L\,+\,l\,L\,+\,L\,l]^{ab}_{ij}(x,y)\, .
\end{eqnarray}
%
%This allows us straightforwardly  to calculate the correlation of the field $\gamma$ eq.(\ref{bdelta},\ref{delta})
%\begin{eqnarray}
%\langle \gamma^a_i(x)\gamma^b_j(y)\rangle&=&{ 4\over 2\pi}\ln{1\over
%x}\left\{D_i [{1\over D\partial}D-{1\over \partial D}\partial]
%[1-l-L+lL+Ll][D{1\over \partial D}-\partial{1\over D\partial}]D_{j}\right\}^{ab}(x,y)\nonumber\\
%&=& { 4\over 2\pi}\ln{1\over x}\left\{D_i[{1\over
%\partial^2}+{1\over D^2}-{1\over \partial^2}\partial D{1\over
%D^2}-{1\over D^2}D\partial{1\over \partial^2}]D_j\right\}^{ab}(x,y)\, .
%\end{eqnarray}
%
Using eq.(\ref{j1}) we then find
\begin{eqnarray}
\chi^{ab}(x,y)&\equiv&{\langle\delta_1 j^a(x)\delta_1 j^b(y)\rangle\over \Delta Y}= 
{4\over \Delta Y}
\left[\partial { D}\frac{1}{ { D}\partial}{ D}_i-\partial_i\right]^{ac}(x,u)\langle c^c_i(0,u)c^d_j(0,v)\rangle
\left[\partial_j-{ D}_j\frac{1}{\partial { D}}{ D}\partial \right]^{db}(v,y)\nonumber\\
&=&
\frac{
1}{\pi}\ \left\{\partial { D}[\frac{1}{
\partial^2}+\frac{1}{ { D}^2}-\frac{1}{\partial^2}\partial { D}\frac{1}{{ D}^2}-\frac{1}{{ D}^2}{ D}\partial\frac{1}{
\partial^2}]{ D}\partial\right\}^{ab}(x,y)\, .
\end{eqnarray}

%%%%%%%%%%%%%%%%%%%%%%%%%%%%%%%%%%%%%%%%%%%%%%%%%
\subsection {From $j$ to $U$.}
To get the evolution equation in the familiar JIMWLK form we need to change variables from $j$ to the single gluon scattering matrix $U$. The matrix $U$ is defined as the matrix of the two dimensional gauge transformation which transforms the 'classical field' $b$ to zero value~\cite{JIMWLK}
\begin{equation}
U^{ab}(x)=\left\{{\cal P}\exp [ig\int_Cdy_i T^cb^c_i(y)]\right\}^{ab}\, ,
\end{equation}
where the contour $C$ starts at some fixed point at infinity in the transverse plane and ends at the point $x$. The matrix $U$ does not depend on the curve $C$ but only on its end point, since the field $b$ is two dimensionally a pure gauge. Using this definition we have
\begin{eqnarray}
\frac{\delta U^{ab}(x)}{\delta j^c(z)} &=&
g\int_Cdy_i\left[U(x)U^\dagger(y)T^d\frac{\delta b^d_i(y)}{ \delta j^c(z)}U(y)\right]^{ab}
\nonumber \\
&=& g\int_Cdy_i\left[U(x)U^\dagger(y)T^dU(y)\right]^{ab}[{ D}_i\frac{1}{\partial { D}}]^{dc}(y,z)\, .
\label{deriv}
\end{eqnarray}
Now we use the identity
\begin{equation}
[U^\dagger(y)T^dU(y)]^{ab}=T^c_{ab}U^{cd}(y)\, .
\end{equation}
Substituting this into eq.(\ref{deriv}), and using the fact that $\int_c dy_i\partial_i F(y)=F(x)$ we find
\begin{equation}
\frac{\delta U^{ab}(x)}{ \delta j^c(z)}=g\left[ UT^b\frac{1}{\partial { D}}\right]^{ac}(x,z)\, .
\end{equation}
This makes it possible to rewrite the real part of the JIMWLK kernel in the following form
\begin{eqnarray}\label{kernel}
\int_{x,y}\chi^{ab}(x,y)\frac{\delta}{\delta j^a(x)}\frac{\delta}{\delta j^b(y)}
&=& \frac{ g^2}{ \pi}\ \int_{x,y}\frac{\delta}{\delta U^{ab}(x)}
\frac{\delta}{\delta U^{cd}(y)}[U(x)T^b]^{al}[U(y)T^d]^{cm}
\nonumber\\
&& \qquad \quad \quad \times
\left[\frac{1}{ \partial^2}+\frac{1}{{ D}^2}-\frac{1}{\partial^2}\partial { D}\frac{1}{ { D}^2}-\frac{1}{ { D}^2}{ D}\partial\frac{1}{\partial^2}\right]^{lm}(x,y)\, .
\end{eqnarray}
Now remember that
\begin{equation}
\frac{\delta}{\delta U^{ab}(x)}[U(x)T^b]^{al}
=-{\rm Tr}\left[\frac{\delta}{\delta U^{\dagger}(x)}U(x)T^l\right]=-J_R^l\, ,
\end{equation}
where $J_R$ is the operator of right rotation on matrix $U$.
We also note that
\begin{equation}
\partial_i\frac{1}{\partial^2}(x,y)=\frac{1}{2\pi}\frac{x_i-y_i}{(x-y)^2};\ \ \ \ \ \ \
{ D}_i\frac{1}{{ D}^2}(x,y)=\frac{1}{2\pi}U^\dagger(x)\frac{x_i-y_i}{(x-y)^2}U(y)\, .
\end{equation}
Now, using eq.(\ref{herm}) we can write the complete kernel as
\begin{equation}
H^{JIMWLK}=-\frac{\alpha_s}{ 2\pi^2}\int_{x,y,z}\frac{(x-z)_i(y-z)_i}{(x-z)^2(y-z)^2}\left[J_L^a(x)J_L^a(y)+J_R^a(x)J_R^a(y)-2J_L^a(x)U^{ab}(z)J_R^b(y)
%-J_R^a(x)U^{ab}(z)J_L^b(y)
\right]
\end{equation}
with $J^a_L(x)=U^{ab}(x)J^b_R(x)$. This is by now one of the standard forms of the JIMWLK kernel, 
see~\cite{remarks}.
%%%%%%%%%%%%%%%%%%%%%%%%%%%%%%%%%%
\subsection{The KLWMIJ evolution}
Although our derivation has been formally in the high density limit, as we noted in the introduction and as we explain in the next section the result eqs.(\ref{J},\ref{j1},\ref{j2}) is in fact valid for all physically interesting situations, including the low density case $j=O(g)$. For the low density case we have to reproduce the  KLWMIJ evolution equation \cite{klwmij},\cite{duality}. It is easy to see that this is indeed the case. Examining the action of the Bogoliubov operator $\cal B$ on the fields, we see that in the weak field limit they are sub leading.  The shift of the vector potential affected by the coherent part of the operator ${\cal C}$ is of order $b\sim j$, while any correction introduced by $\cal B$ is of order $gb\sim gj$.  This is also true in the strong field case, however for $j\sim 1/g$ the corrections due to $\cal B$ are $O(1)$ and therefore could not be neglected. In the weak field case these are not only sub leading but also genuinely perturbative!
 We can therefore neglect the action of 
$\cal B$ altogether. Thus in this limit the operator $\Omega$ reduces to the coherent operator ${\cal C}$ with the 'classical field' $b$ given by the leading order perturbative expression. This is precisely the operator that was used in \cite{klwmij} to derive KLWMIJ evolution equation. Obviously, repeating the same derivation we obtain the same result.

 One important thing to be noted here is, that in order to derive KLWMIJ we are not allowed to expand the correlators of the transformed charge density to first order in $\delta_2j$ eq.(\ref{j2}) as is done to derive JIMWLK equation. The reason is very simple. When $j\sim O(g)$, the second term on the RHS of eq.(\ref{j2}) is of the same order as $j$ itself . Therefore its contribution to the evolved correlators of $J$ has to be resumed to all orders. This is indeed what is done in the derivation of \cite{klwmij}. It is the resummation to all orders in $ f^{abc}\int dx^-\tilde A^b(x)\partial^+\tilde A^c(x)$ that is responsible for the appearance of the 'dual Wilson line factor' 
$$R(x)\equiv \exp\left[ T^a\frac{\delta}{\delta j^a(x)}\right]$$
 in the KLWMIJ evolution equation \cite{klwmij}.

%%%%%%%%%%%%%%%%%%%%%%%%%%%%%%%%%%%%%%%%%%%%%%%
\section{Discussion}
In this paper we have carried through the diagonalization of the QCD light cone Hamiltonian in the presence of a
 valence charge density $j$. We found that for large valence charge density, to $O(1)$ the vacuum is the Bogoliubov transform 
of the free gluon vacuum. We have also found the action of the Bogoliubov operator on the dynamical variables of QCD 
including the valence color charge density. 
The evolution of hadronic wave function to high energy increases the longitudinal momentum of the gluons 
in this state. Thus more energetic gluons scatter on the target leading to the evolution of the hadronic 
scattering matrix. 
\subsection{The calculation is valid for any $j$.}
We have shown 
that when the valence charge density is large $j\sim O(1/g)$ the wave
function we found leads to the JIMWLK evolution equation. 
However, our calculation itself is valid
beyond the high density limit, and does in fact give the leading
solution of the light cone Hamiltonian  for all physically interesting
magnitudes of the color charge density $j\sim O(g^n)$; $-1\leq n\leq 1$.  The
precise statement is that relative corrections to the solution we have
given here are proportional to a positive power of $g$ at any
interesting value of the valence charge density.

To see this, recall that the basis of our approach was the perturbative solution of eqs.(\ref{constrmag},\ref{constrel1}). 
We have solved eq.(\ref{constrmag}) exactly, while eq.(\ref{constrel1}) was solved treating the second, 
third and fourth terms on the left hand side (LHS) as perturbations.
The solution of this pair of equations to leading order in the coupling constant is always of order 
$\gamma=O(j)=O(g^n)$ for $-1\leq n\leq 1$ . 
The magnitude of corrections is easy to estimate. Since, by definition the field $c$ is $O(1)$, we have
\begin{eqnarray}
&&gf^{abc}[\gamma_i^b(x),\gamma_i^c(x)]\sim g\left(\frac{\delta\gamma}{\delta j}\right)^2[j,j]\sim g^2j=O(g^{2+n})\, ,\nonumber\\
&&gf^{abc}\{\gamma_i^b(x),c^c_i(x,0)\}=O(g^{1+n})\, ,\nonumber\\
&&gf^{abc}\int dx^-c_i^b(x^-)\partial^+c^c_i(x^-)=O(g)\, .
\end{eqnarray}
The first term is always smaller than the second. It always scales as a positive power of $g$ and therefore can always 
be treated perturbatively.

 The second term is also small as long as $n\ne -1$. It is a factor $g$ smaller than the zeroth order solution and 
thus again can be safely treated perturbatively. For $n\ne-1$ it can be neglected since its magnitude is a 
positive power of $g$. The case $n=-1$ is a bit different, since then this term is $O(1)$ and so has to be taken 
into account, which is what we did above. 
 
 Finally the third term is always $O(g)$. It can be neglected for all $n\ne 1$.
%As for eq.(\ref{constrel1}), since by definition $c\sim O(1)$, the second and third terms of the left hand side are always sub leading and so is the fourth term as long as $n\ne 1$. Thus as long as $n\ne 1$ and $n\ne -1$, the leading term in the solution is $O(g^n)$ while the correction is $O(g^{1+n})<1$ and can be neglected. For $n=-1$ the correction is itself $O(1)$, and has to be taken into account which we did in this paper. 
For $n=1$ this term is of the same magnitude as $j$ and thus it may seem that it has to be taken into account already in the leading order. However this is not the case for the following reason. The vacuum of the Hamiltonian of the field $c$ at $j\sim O(g)$ is a free vacuum. This state is annihilated by the 'soft' color charge density operator
$f^{abc}\int dx^-c^b(x)\partial^+c^c(x)$. Thus this operator only gives non vanishing contribution to $\gamma$ in the sub leading order in $g$, where the vacuum is not a free vacuum anymore. Thus we see that for all $-1\leq n\leq 1$ our solution of eqs.(\ref{constrmag},\ref{constrel1}) keeps the leading terms and for $n=-1$ also the important sub leading term of $O(1)$. The terms that we omit are not only suppressed by a positive power of $g$ relative to the terms we keep, but also vanish in the limit $g\rightarrow 0$ at any $j$. 

To reiterate, our procedure keeps all the terms that are important for physically interesting values of the color charge density. This is not to say that our solution can be considered as a leading order of some expansion which has the same expansion parameter for all $n$. The corrections to the leading term may  have different magnitude for different values of $n$, and thus the properties of the expansion are different at different values of $n$. At this point however we are not interested in the sub dominant corrections and will not discuss this issue any further.

\subsection{What JIMWLK misses?}
We want now to return to the point briefly mentioned at the end of the
previous section. Even though our diagonalization procedure and the solution for the vacuum wave function is valid
for any $j$, the derivation of the evolution equation {\it for the
scattering amplitude} involves one extra step, and that is adding  the
charge density of the soft gluons to the valence charge density. For
$n\ne 1$ this is a perturbative proposition, since the soft gluon
charge density  is parametrically smaller than $j$ itself. Thus for
the derivation of the JIMWLK evolution equation one expands to first
order in the soft gluon color charge density, the second term on the
RHS of eq.(\ref{j2}). For the KLWMIJ evolution on the other hand all powers of the soft gluon color charge density are resummed.
The addition  of the soft gluon charge density is achieved by acting on
any observable function of $j$ by the shift operator of the form
\begin{equation}\label{r}
\hat R_a\,=\,\exp\left[ \int d^2x\,j_{soft}^a(x)\,\frac{\delta}{\delta j^a(x)}\right] \, .
\end{equation}
In the KLWMIJ limit only one gluon is produced at one step of the
evolution with probability of order $\alpha_s$, and thus $j_{soft}^a(x)=T^a$ when acting on the component
of the wave function which contains this extra gluon. The phase
factor of eq.(\ref{r}) therefore simply becomes the dual Wilson loop $R$. 
In the general case however the action of the Bogoliubov operator
$\cal B$ produces an arbitrary number of gluons.  For $j\sim 1/g$ the number of gluons of order $O(1)$ is
produced with probability of $O(1)$, while with probability
$O(g)$ one can produce $O(1/g)$ extra
gluons. The phase factor becomes a product of dual Wilson loops
$R(x_1)...R(x_n)$ when acting on a component of the wave function with
$n$ extra gluons.
Now the JIMWLK equation is valid when a large dense target scatters
off a small perturbative projectile. In this situation each gluon in
the target wave function undergoes only a small number of scatterings on the
projectile. In fact the leading order scattering on a small target is only due to two gluon exchange.
This corresponds to expansion of each dual Wilson loop
factor $R$ to second order in $\delta/\delta j$. 
It is also true that in this situation it is unlikely that two or more produced gluons scatter simultaneously. 
Indeed the expansion of eq.(\ref{r}) is equivalent to approximating the scattering amplitude of the configuration 
of $n$ produced gluons by the sum of the individual scattering amplitudes.  

Recall that in calculating the evolution of any correlation
 function of $j$ in the JIMWLK approximation, we only keep terms of the first order in $\delta_2j$ and of the
 second order in $\delta_1j$. Thus the correction to the scattering matrix 
$S=\exp \{i\,j\,\alpha_T\}$ due to the evolution is
 at most of second order in the target field $\alpha_T$. This is another way of saying that the whole system of soft 
gluons produced in one step of the evolution scatters 
on the target only  via the two gluon exchange. To be a little more  precise we have to remember that while calculating
the evolution of the scattering matrix, a 
factor $R$
accompanies not only each soft emitted gluon but also every factor of $j$ in the operator $\Omega$. 
The eikonal scattering matrix of the projectile wave function on the target field $\alpha_T$ is given by (we drop the transverse coordinate dependence to simplify the notations) 
\begin{eqnarray}\label{ampt}
\Sigma^P&=&\langle \Psi[j]|\Omega^\dagger[j,a,a^\dagger]e^{i(j^a+j_{soft}^a)\alpha_T^a}\Omega[j,a,a^\dagger]|\Psi[j]\rangle=
\langle \Psi[j]|\Omega^\dagger[j,a]e^{ij^a\alpha^a_T} \Omega[j,Ra,Ra^\dagger]|\Psi[j]\rangle\nonumber\\&=&
\langle \Psi[j]|\Omega^\dagger[j,a]\Omega[Rj,Ra, Ra^\dagger] e^{ij^a\alpha_T^a} |\Psi[j]\rangle\, ,
\end{eqnarray}
where $|\Psi[j]\rangle$ is the valence wave function and the functional derivatives in $R$ act only on the eikonal factor $e^{i\,j^a\,\alpha_T^a}$. The first equality is the reflection of the fact that multiplying every soft gluon creation operator by $R$ is equivalent to shifting the charge density $j$ by the charge density of this soft gluon. The second equality follows from commuting of the operators $j$ in $\Omega$ with those in the eikonal factor as explained in detail in \cite{R}. Since every  $R$ becomes an eikonal factor after acting on $e^{i\,j^a\,\alpha_T^a}$, multiplication of $j$ by $R$ in the second line in eq.(\ref{ampt}) physically corresponds to the effect of scattering of the valence charges involved in the emission of soft gluons.
Thus the expansion
 of all the factors
of $R$ to second order in $\frac{\delta}{\delta j}$ approximates the interaction of the whole system of soft gluons 
emitted in 
one step of the evolution plus the valence charges involved in their emission (in the following we will refer to this system as "soft gluons" to avoid lengthy and wordy descriptions), with the target by a two gluon exchange.

The JIMWLK evolution therefore does not take into account multiple scattering corrections to the amplitude due to simultaneous scattering of two soft gluons emitted in the same step of the evolution. This is not to say that the JIMWLK evolution does not allow any multiple scattering corrections at all. In particular the probability that a soft gluon scatters simultaneously with some of the valence gluons not participating in its emission, is accounted for. We will refer to these multiple scattering events as "long range multiple scatterings" to emphasize the fact that the two objects that scatter simultaneously have vastly different rapidities. This as opposed to "short range multiple scatterings" where both objects have similar rapidity, which are taken into account by the KLWMIJ evolution. 

It is these long range multiple scattering corrections that unitarize the scattering amplitude in the JIMWLK approximation. If no multiple scattering corrections where included at all, the amplitude would not unitarize even though the coherent effects in the wave function are taken into account exactly. Recall that the charge density itself does not saturate even in the dense regime, although its growth with rapidity is much slower than in the BFKL approximation \cite{zakopane}. 
In particular in the BFKL (or equivalently KLWMIJ) limit the color charge density grows exponentially with rapidity
\begin{equation}\label{jdilute}
j^2(Y)\propto j(0)^2e^{\omega Y}\, ,
\end{equation}
while in the "saturated regime" the growth is a random walk process and thus \cite{zakopane}
\begin{equation}\label{jdense}
j^2\propto j(0)^2+kY\, .
\end{equation}
Since the charge density does not stop growing even in the saturated regime, the scattering amplitude would not saturate if no multiple scattering corrections are taken into account. It is thus precisely the long range multiple scattering corrections that stop the scattering  amplitude from growing beyond one in the JIMWLK approximation.

Eqs.(\ref{jdilute}) and (\ref{jdense}) in fact clearly indicate that the short range  multiple scatterings are dominant in the KLWMIJ  regime while the long range multiple scatterings are dominant in the JIMWLK regime. Consider first the evolution of a dilute projectile (KLWMIJ evolution). According to eq.(\ref{jdilute}) the color charge density grows exponentially fast and is always (at large enough rapidity) dominated by gluons  created in the last rapidity interval of the size $\Delta Y\approx \frac{1}{\omega}$. Thus the dominant multiple scattering effects indeed are due to the simultaneous scattering of two or more gluons at approximately the same rapidity - the "short range multiple scatterings". On the other hand in the JIMWLK regime where eq.(\ref{jdense}) is valid, the color charge density is uniformly distributed in rapidity. Thus clearly the dominant multiple scattering corrections are due to simultaneous scatterings of gluons at far away rapidities - the "`long 
range multiple scatterings"'. 

We thus see explicitly that while the KLWMIJ evolution takes into account all multiple scattering effects but does 
not include nonlinearities in the evolution of the wave function, the JIMWLK evolution fails to take account 
of the short range multiple scattering corrections to the amplitude.
\subsection{Short range multiple scattering and the dipole - dipole amplitude.}
In relation to the preceding discussion we want to comment briefly on one aspect of the
Pomeron loop correction to the JIMWLK evolution. In particular recently much attention has been devoted to scattering of two 
unequal size dipoles. In this context there has been much discussion of the effects of discreteness and fluctuations in the target (taken to be the larger of the two dipoles) wave function\cite{stat}. Although our derivation does not indicate any reason to expect that 
discreteness and/or fluctuations are particularly important, it does indeed show that the application of the JIMWLK or KLWMIJ evolution to the target wave function in the dipole-dipole scattering  is flawed. 
The reason KLWMIJ evolution fails is obvious. Starting with a dilute single dipole target  initial stages of the evolution are indeed well described by the KLWMIJ equation. However when the density in the target wave function reaches large value $j\propto 1/g$ neglecting high density effects in the evolution of the wave function is not permissible. This density is parametrically the same as that for which the scattering amplitude becomes of order one, and it is therefore also the same density at which  the effect of the multiple scattering corrections in KLWMIJ evolution becomes significant. This has been recognized in the literature for a long time, see for example fifth paper in \cite{JIMWLK}. 

On the other hand the reason for the failure of JIMWLK is somewhat more subtle. Again starting with the dilute target one can initially evolve it with the JIMWLK equation. The multiple scattering effects are not important as long as the density is small, and thus the use of JIMWLK in the dilute regime is as good as the use of KLWMIJ. When the density is parametrically large again the JIMWLK evolution is valid, since the evolution of the wave function is accounted for appropriately and the long range multiple scattering corrections dominate at high density. It might therefore seem that JIMWLK equation can be used all the way through in this situation. This is however not the case. The reason it fails is that there is a range of rapidities in the evolution when the density is already not very small but the rate of growth is still large. This happens just before the 
saturation is reached. Since the density in this range of rapidities still grows exponentially, the short range multiple scattering effects dominate. Those are not included in JIMWLK evolution, and thus the rate of growth of the amplitude is overestimated. Note that if already at the initial rapidity the density in the target wave function is large (e.g. for a heavy nucleus) there is no rapidity window in which the short range multiple scatterings dominate, and thus JIMWLK evolution is valid.

We close the discussion by stressing that the calculation of the wave function given in the present paper is the correct starting point for derivation of 
the complete evolution equation which takes into account 
all relevant Pomeron loop effects. The validity of such equation will not be limited to the process of collision of two small objects, but more interestingly to the situation where two colliding objects are large. The use of JIMWLK evolution in this case is not justified since the soft gluons produced in the wave function can multiply rescatter on the large target field.

\medskip
\section*{Acknowledgments}
\medskip

We thank the Institute for Nuclear Theory at the University of Washington for its hospitality and the Department
of Energy for partial support during the completion of this work. 
 
A.K. thanks the Galileo Galilei Institute of the University of Florence for hospitality and financial support and participants of the GGI program "High density QCD" for informative discussions while part of this work was being done.
The work of A.K. is supported by the US-DOE grant DE-FG02-92ER40716.

The work of M.L. is  supported by the US-DOE grants DE-FG02-88ER40388
and DE-FG03-97ER4014.

%%%%%%%%%%%%%%%%%%%%%%%%%%%%%%%%%%%%%%%%%%%%%%%%%%%%%%%%%%%%%%%%%%%%%%%%%%%%%%%%

%%%%%%%%%%%%%%%%%%%%%%%%%%%%%%%%%%%%%%%%%%%%%%%%%%%%%%%%%%%%%%%%%%%%%%%%%%%%%%%%

\end{document}